# Information-theoretic Limits on Learning and Estimation

**Abbas El Gamal* and Maxim Raginsky****

Stanford University*, UIUC**

# Limits on Learning and Estimation

Information theory plays a central role in establishing fundamental limits on what any learning or estimation algorithm can—and cannot—achieve, regardless of computational power. In this chapter, we present several key ideas and results that demonstrate this connection.

We begin by introducing the tools needed for the analysis. First, we derive classical concentration inequalities, which provide probabilistic bounds on how much empirical quantities—such as sample means—can deviate from their expected values. We then discuss covering and packing in metric spaces and define metric entropy, which quantifies the complexity of a function class as the logarithm of the minimum number of balls of a given radius required to cover it. These geometric concepts are closely related to the probabilistic covering and packing ideas introduced earlier in the study of source and channel coding.

Using these tools, we address two conceptually distinct yet mathematically related topics. The first concerns upper bounds on generalization error in statistical learning, which quantify how well a learning algorithm performs on unseen data. The second concerns lower bounds on minimax estimation risk, which characterize the intrinsic difficulty of estimation problems.

After describing the learning-theoretic setup, we derive upper bounds on the generalization error in terms of measures of model class complexity—most notably metric entropy, Rademacher complexity, and the VC dimension—as well as mutual information and relative entropy. In many standard settings, the resulting bounds scale on the order of $O\left(C/\sqrt{n}\,\right)$, where $C$ denotes the model complexity (capacity) and $n$ is the sample size.

We then introduce the minimax estimation framework. Since computing the minimax risk exactly is generally intractable, we focus on establishing lower bounds. We first reduce the estimation problem to a multiple-hypothesis testing problem over a suitably constructed finite packing of the parameter space. Applying Fano's inequality relates the probability of error in this testing problem to the mutual information between the observations and the underlying hypothesis, yielding fundamental limits on the achievable performance of any estimator. For many classical problems, such as Gaussian mean estimation, these bounds are order-optimal.

## 0.1 CONCENTRATION INEQUALITIES

The term "concentration of measure" refers to the phenomenon that a function of many independent or weakly dependent random variables is nearly constant with high probability, typically having small fluctuations around its expected value. Concentration inequalities provide bounds on the difference between the function and its expected value. As an illustration, consider the sample average

$$\phi(X_1, X_2, \ldots, X_n) = \frac{1}{n}\sum_{i=1}^{n} X_i, \tag{0.1}$$

where $X_1, X_2, \ldots, X_n$ are i.i.d. random variables with finite mean $\mathsf{E}(X_1) = \mu$ and variance $\sigma^2 = \mathrm{Var}(X_1)$. Letting $U = \phi(X^n)$, we have $\mathsf{E}(U) = \mu$ and $\mathrm{Var}(U) = \sigma^2/n$.

As a first attempt to bound the probability that $U$ deviates from $\mu$ by more than a prescribed amount $\epsilon > 0$, we can apply Chebyshev's inequality to obtain

$$\mathsf{Pr}\{|U - \mu| \geq \epsilon\} \leq \frac{\sigma^2}{n\epsilon^2}.$$

How sharp is this bound?

If all we know about a random variable is its mean and variance, then the Chebyshev inequality cannot, in general, be improved. However, we can obtain substantially tighter bounds if we have more information about the distribution of the random variable.

For example, consider the case in which the $X_i$'s are i.i.d. zero-mean Gaussian random variables. Here the mean and variance are finite, but we can obtain a significantly sharper bound than that provided by Chebyshev's inequality using the closed form expression for the moment generating function of $X \sim \mathrm{N}(0, \sigma^2)$,

$$\mathsf{E}\left(e^{\lambda X}\right) = e^{\lambda^2\sigma^2/2}. \tag{0.2}$$

Following the derivation of the Chernoff bound, we obtain, for any $\lambda > 0$,

$$\begin{aligned}
\mathsf{Pr}\{X \geq \epsilon\} &= \mathsf{Pr}\{e^{\lambda X} \geq e^{\lambda\epsilon}\} && (0.3)\\
&\leq e^{-\lambda\epsilon}\,\mathsf{E}\left(e^{\lambda X}\right) && (0.4)\\
&= \exp\Big(-\lambda\epsilon + \frac{\lambda^2\sigma^2}{2}\Big). && (0.5)
\end{aligned}$$

Optimizing over $\lambda$ yields

$$\begin{aligned}
\mathsf{Pr}\{X \geq \epsilon\} &\leq \inf_{\lambda>0} \exp\Big(-\lambda\epsilon + \frac{\lambda^2\sigma^2}{2}\Big) && (0.6)\\
&= \exp\Big(-\frac{\epsilon^2}{2\sigma^2}\Big). && (0.7)
\end{aligned}$$

Returning to the sample average $U = \frac{1}{n}\sum_{i=1}^{n} X_i$, where the $X_i$'s are i.i.d. $\mathrm{N}(0, \sigma^2)$ random variables, we note that $U \sim \mathrm{N}(0, \sigma^2/n)$. Applying the preceding bound with variance $\sigma^2/n$, we obtain

$$\Pr\{U \geq \epsilon\} \leq \exp\Big(-\frac{n\epsilon^2}{2\sigma^2}\Big). \tag{0.8}$$

Applying the same arguments to $-U$ and using the union bound yields the two-sided tail bound

$$\Pr\{|U| \geq \epsilon\} \leq 2\exp\Big(-\frac{n\epsilon^2}{2\sigma^2}\Big), \tag{0.9}$$

which decays exponentially in $n$, in contrast to the polynomial decay provided by Chebyshev's inequality.

**Remark.** In the Gaussian case, the tail behavior can be characterized more precisely, leading to the sharper bound (see Problem 0.1)

$$\Pr\{|U| \geq \epsilon\} \leq \frac{2\sigma}{\epsilon\sqrt{2\pi n}}\exp\Big(-\frac{n\epsilon^2}{2\sigma^2}\Big). \tag{0.10}$$

### 0.1.1 Subgaussian distributions and Hoeffding's inequality

The preceding discussion motivates the definition of the following important class of random variables.

**Definition 0.1.** A random variable $X$ is said to be $\sigma^2$-*subgaussian* if there exists a finite constant $\sigma^2 > 0$ such that

$$\mathsf{E}\Big(e^{\lambda(X-\mathsf{E}(X))}\Big) \leq e^{\lambda^2\sigma^2/2} \quad \text{for all } \lambda \in \mathbb{R}. \tag{0.11}$$

The parameter $\sigma^2$ is often referred to as the *variance proxy*. Taking the logarithm of (0.11), we obtain the equivalent definition: $X$ is $\sigma^2$-subgaussian if

$$\ln \mathsf{E}\Big(e^{\lambda X}\Big) \leq \lambda\,\mathsf{E}(X) + \frac{\lambda^2\sigma^2}{2} \quad \text{for all } \lambda \in \mathbb{R}. \tag{0.12}$$

**Example 0.1.** Let $X \sim \mathrm{N}(\mu, \sigma^2)$. Then, the bound in (0.11) is achieved with equality. Hence, $X$ is $\sigma^2$-subgaussian.

A discrete random variable can also be subgaussian. Consider the example.

**Example 0.2.** The *Rademacher random variable*, usually denoted by $\varepsilon$, takes values in $\{-1, 1\}$ with equal probability. To see that it is 1-subgaussian, consider

$$\mathsf{E}\Big(e^{\lambda\varepsilon}\Big) = \frac{1}{2}e^{-\lambda} + \frac{1}{2}e^{\lambda} \tag{0.13}$$

$$= \frac{1}{2}\sum_{k=0}^{\infty}\frac{(-\lambda)^k}{k!} + \frac{1}{2}\sum_{k=0}^{\infty}\frac{\lambda^k}{k!} \tag{0.14}$$

$$= \sum_{k=0}^{\infty} \frac{(\lambda^2)^k}{(2k)!} \tag{0.15}$$

$$\leq \sum_{k=0}^{\infty} \frac{(\lambda^2)^k}{(2^k) \cdot k!} \tag{0.16}$$

$$= \exp\Big(\frac{\lambda^2}{2}\Big). \tag{0.17}$$

A more general result states that every bounded random variable is subgaussian.

**Lemma 0.1** (Hoeffding)**.** Let $X$ be a random variable taking values in $[a, b]$ for some $-\infty < a \leq b < \infty$. Then $X$ is $\sigma^2$-subgaussian with $\sigma^2 = (b-a)^2/4$.

**Proof.** Without loss of generality, we may assume that $\mathsf{E}(X) = 0$. Define the cumulant generating function $\psi(\lambda) = \log \mathsf{E}\left(e^{\lambda X}\right)$. Differentiating twice with respect to $\lambda$ (and noting that we can exchange the derivative with expectation using the boundedness of $X$ and the dominated convergence theorem) gives

$$\psi'(\lambda) = \frac{\mathsf{E}\left(Xe^{\lambda X}\right)}{\mathsf{E}\left(e^{\lambda X}\right)}, \tag{0.18}$$

$$\psi''(\lambda) = \frac{\mathsf{E}\left(X^2 e^{\lambda X}\right)}{\mathsf{E}\left(e^{\lambda X}\right)} - \left(\frac{\mathsf{E}\left(Xe^{\lambda X}\right)}{\mathsf{E}\left(e^{\lambda X}\right)}\right)^2. \tag{0.19}$$

For each $\lambda \in \mathbb{R}$, define the *tilted* random variable $Z_\lambda$ by the distribution

$$p_{Z_\lambda}(x) = \frac{e^{\lambda x}}{\mathsf{E}\left(e^{\lambda X}\right)} p_X(x). \tag{0.20}$$

With this definition, $\psi''(\lambda) = \mathrm{Var}(Z_\lambda)$. Moreover, since $X$ takes values in the interval $[a, b]$, so does $Z_\lambda$, which implies that

$$\psi''(\lambda) = \mathrm{Var}(Z_\lambda) \leq \frac{(b-a)^2}{4}. \tag{0.21}$$

The proof of this inequality is left as an exercise (Problem 0.4).

Using the fact that $\psi(0) = \psi'(0) = 0$ and the fundamental theorem of calculus, we have

$$\psi(\lambda) = \int_0^{\lambda} \int_0^{\lambda'} \psi''(\tau)\, d\tau d\lambda' \leq \frac{\lambda^2 (b-a)^2}{8}. \tag{0.22}$$

Exponentiating both sides then gives

$$\mathsf{E}\left(e^{\lambda X}\right) \leq \exp\Big(\frac{\lambda^2(b-a)^2}{8}\Big),$$

which completes the proof of the lemma.

Following the same steps ((0.3)–(0.7)) as for the Gaussian case, we arrive at the following key result.

**Lemma 0.2** (Subgaussian tail bound)**.** Let $X$ be a $\sigma^2$-subgaussian random variable. Then, for any $\epsilon > 0$,

$$\mathsf{Pr}\{X - \mathsf{E}(X) \geq \epsilon\} \leq \exp\Big(-\frac{\epsilon^2}{2\sigma^2}\Big), \tag{0.23}$$

$$\mathsf{Pr}\{X - \mathsf{E}(X) \leq -\epsilon\} \leq \exp\Big(-\frac{\epsilon^2}{2\sigma^2}\Big). \tag{0.24}$$

Obtaining concentration inequalities in certain settings (see, e.g., Theorem 0.10) requires bounding the moment generating function of the square of a subgaussian random variable.

**Lemma 0.3.** Let $X$ be a zero-mean $\sigma^2$-subgaussian random variable. For any $0 \leq \lambda < 1/2\sigma^2$,

$$\mathsf{E}\left(e^{\lambda X^2}\right) \leq \frac{1}{\sqrt{1 - 2\lambda\sigma^2}}.$$

**Proof.** The lemma holds if $X$ is Gaussian (see Problem 0.3). To extend it to an arbitrary subgaussian random variable, we introduce a *ghost* (*auxiliary*) random variable $Z \sim \mathrm{N}(0, 1)$ independent of $X$, and apply the subgaussian condition for each outcome of $z$. Using the moment generating function of $Z$, we can write

$$e^{\lambda X^2} = \mathsf{E}_Z\left(e^{\sqrt{2\lambda} XZ} \middle| X\right).$$

Taking expectation with respect to $X$ and interchanging expectation yields

$$\mathsf{E}\left(e^{\lambda X^2}\right) = \mathsf{E}_X\left(\mathsf{E}_Z\left(e^{\sqrt{2\lambda} XZ} \middle| X\right)\right) \tag{0.25}$$

$$= \mathsf{E}_Z\left(\mathsf{E}_X\left(e^{\sqrt{2\lambda} XZ}\right)\right) \tag{0.26}$$

$$\leq \mathsf{E}_Z\left(e^{\lambda Z^2 \sigma^2}\right) \tag{0.27}$$

$$= \frac{1}{\sqrt{1 - 2\lambda\sigma^2}}, \tag{0.28}$$

where (0.27) follows from the fact that, for each fixed $z$, the random variable $zX$ is $(z\sigma)^2$-subgaussian.

### 0.1.2 Maximal inequalities

As an application of the above results, we analyze the behavior of the *maximum* of finitely many, not necessarily independent, subgaussian random variables. This analysis will be useful later in the context of learning. Let $X_1, X_2, \ldots, X_n$ be zero-mean, $\sigma^2$-subgaussian random variables. We begin by deriving an upper bound on the expected value of their maximum.

**Lemma 0.4** (Maximal Inequality for Subgaussians)**.** Let $X_1, X_2, \ldots, X_n$, $n \geq 2$, be zero-mean $\sigma^2$-subgaussian random variables. Then

$$\mathsf{E}\Big(\max_{1\leq i\leq n} X_i\Big) \leq \sqrt{2\sigma^2 \ln n}. \tag{0.29}$$

**Proof.** Fix $\lambda > 0$ and consider

$$\begin{aligned}\mathsf{E}\Big(\max_{1\leq i\leq n} X_i\Big) &= \frac{1}{\lambda}\mathsf{E}\Big(\ln\exp\big(\lambda\max_{1\leq i\leq n} X_i\big)\Big) &(0.30)\\ &\leq \frac{1}{\lambda}\ln\mathsf{E}\Big(\exp\big(\lambda\max_{1\leq i\leq n} X_i\big)\Big) &(0.31)\\ &= \frac{1}{\lambda}\ln\mathsf{E}\Big(\max_{1\leq i\leq n} e^{\lambda X_i}\Big) &(0.32)\\ &\leq \frac{1}{\lambda}\ln\sum_{i=1}^n \mathsf{E}\big(e^{\lambda X_i}\big), &(0.33)\end{aligned}$$

where (0.31) follows from Jensen's inequality, (0.32) follows since the exponential function is monotonically increasing, and the last step follows from the bound $\max_i a_i \leq \sum_i a_i$ for nonnegative $a_i$s. Since each $X_i$ is $\sigma^2$-subgaussian and has zero mean,

$$\mathsf{E}\Big(e^{\lambda X_i}\Big) \leq \exp\Big(\frac{\lambda^2\sigma^2}{2}\Big), \tag{0.34}$$

therefore

$$\mathsf{E}\big(\max_{1\leq i\leq n} X_i\big) \leq \frac{\ln n}{\lambda} + \frac{\lambda\sigma^2}{2}. \tag{0.35}$$

Since the left-hand side does not depend on $\lambda$, minimizing the right-hand side yields the choice $\lambda = \sqrt{2\ln n/\sigma^2}$. This establishes the lemma.

**Remark.** It is not difficult to show (Problem 0.8) using essentially the same method that

$$\mathsf{E}\Big(\max_{1\leq i\leq n} |X_i|\Big) \leq \sqrt{2\sigma^2 \ln(2n)}, \tag{0.36}$$

Note that the independence of the $X_i$s in the above lemma is *not* required. Moreover, we can obtain the following bound.

**Lemma 0.5** (Maximal tail inequality for subgaussians)**.** Let $X_1, X_2, \ldots, X_n$, $n \geq 2$, be zero-mean, $\sigma^2$-subgaussian random variables. Then, for any $\epsilon > 0$,

$$\mathsf{Pr}\Big\{\max_{1\leq i\leq n} X_i \geq \sqrt{2\sigma^2\ln n} + \epsilon\Big\} \leq \exp\Big(-\frac{\epsilon^2}{2\sigma^2}\Big). \tag{0.37}$$

**Proof.** By the union bound and Lemma 0.2, we have

$$\mathsf{Pr}\Big\{\max_{1\leq i\leq n} X_i \geq \sqrt{2\sigma^2\ln n} + \epsilon\Big\} \leq \sum_{i=1}^n \mathsf{Pr}\big\{X_i \geq \sqrt{2\sigma^2\ln n} + \epsilon\big\} \tag{0.38}$$

$$\leq n \exp\left( -\frac{\left(\sqrt{2\sigma^2 \ln n} + \epsilon\right)^2}{2\sigma^2} \right) \tag{0.39}$$

$$= \exp\left( \ln n - \frac{\left(\sqrt{2\sigma^2 \ln n} + \epsilon\right)^2}{2\sigma^2} \right) \tag{0.40}$$

$$\leq \exp\left( -\frac{\epsilon^2}{2\sigma^2} \right), \tag{0.41}$$

which completes the proof.

## 0.2 METRIC ENTROPY

The concept of metric entropy was introduced by Kolmogorov and Tikhomirov (1959) under the name $\epsilon$-entropy. It can be viewed as a deterministic analog of the rate distortion function and can be related to it. We begin with some definitions.

**Definition 0.2.** A *metric space* $(\mathscr{V}, \rho)$ consists of a set of elements $\mathscr{V}$ together with a metric $\rho : \mathscr{V} \times \mathscr{V} \to \mathbb{R}_+$, i.e., a function that is nonnegative, symmetric, satisfies the triangle inequality, and is zero if and only if $v = v'$, such as the Euclidean and Hamming distances.

**Definition 0.3.** Let $\mathscr{S} \subseteq \mathscr{V}$ and $\delta > 0$. A set $\mathcal{V} = \{v_1, v_2, \dots, v_m\} \subset \mathscr{V}$ is said to be a $\delta$-*net* of $\mathscr{S}$ if for every $v \in \mathscr{S}$, $\min_{1 \leq i \leq m} \rho(v, v_i) \leq \delta$.

A $\delta$-net induces a *covering* of $\mathscr{S}$ by balls $B_i(v_i, \delta) = \{v \in \mathscr{S} : \rho(v, v_i) \leq \delta\}$ centered at $v_i \in \mathcal{V}$ for $i = 1, 2, \dots, m$, such that $\mathscr{S} \subseteq \cup_{i=1}^{m} B_i$.

**Definition 0.4.** The *covering number* of $\mathscr{S}$ at resolution $\delta$, denoted by $N(\delta; \mathscr{S}, \rho)$, is the size of the smallest $\delta$-net for $\mathscr{S}$, that is, $N(\delta; \mathscr{S}, \rho) = \min\{|\mathcal{V}| : \mathcal{V} \subseteq \mathscr{V} \text{ is a } \delta\text{-net for } \mathscr{S}\}$.

**Definition 0.5.** The *metric entropy* of $\mathscr{S}$ at resolution $\delta$ is the logarithm of the covering number

$$H(\delta; \mathscr{S}, \rho) = \log N(\delta; \mathscr{S}, \rho). \tag{0.42}$$

When $\mathscr{S}$ is such that $H(\delta; \mathscr{S}, \rho) < \infty$ for all $\delta > 0$, we say that $\mathscr{S}$ is *totally bounded*.

Consider the following examples.

**Example 0.3** (Euclidean Ball)**.** Let $B(0, r) = \{\mathbf{x} \in \mathbb{R}^n : \|\mathbf{x}\|_2 \leq r\}$ (or $B(r)$ in short) denote the closed Euclidean ball of radius $r$ centered at the origin, and let $\rho$ be the Euclidean distance. Then the metric entropy of $B(0, r)$ at resolution $\delta > 0$ is bounded as

$$n \log\left(\frac{r}{\delta}\right) \leq H(\delta; B(r), \rho) \leq n \log\left(1 + \frac{2r}{\delta}\right). \tag{0.43}$$

The lower bound follows from a simple volume argument. For the upper bound, let $\mathcal{V} = \{\mathbf{v}_1, \mathbf{v}_2, \dots, \mathbf{v}_m\} \subset B(r)$ be a set of maximum cardinality such that $\|\mathbf{v}_i - \mathbf{v}_j\|_2 > \delta$ for every $i \neq j$. Then we claim that $\mathcal{V}$ forms a $\delta$-net of $B(r)$. Suppose this is not the case, then there exists a point $\mathbf{v} \in B(r)$ such that $\|\mathbf{v} - \mathbf{v}_i\|_2 > \delta$ for every $i$. But then we could add $\mathbf{v}$ to $\mathcal{V}$, obtaining a strictly larger set whose elements still have pairwise distances greater than $\delta$. This contradicts the maximality of $\mathcal{V}$. Thus $\mathcal{V}$ must form a $\delta$-net of $B(r)$. By definition, the closed balls of radius $\delta/2$ centered on the points of $\mathcal{V}$ are disjoint. Moreover, they all lie within the ball $B(r+\delta/2)$. Hence, $m\,\mathrm{Vol}(B(\delta/2)) \le \mathrm{Vol}(B(r+\delta/2))$, and the upper bound in (0.43) follows.

**Example 0.4** (Hamming cube)**.** Consider the metric space consisting of the set of $n$-bit sequences $\{0,1\}^n$ with the Hamming metric $\rho_{\mathrm{H}}(\mathbf{x},\mathbf{x}') = \sum_{i=1}^n \mathbb{1}_{\{x_i \neq x_i'\}}$, $\mathbf{x},\mathbf{x}' \in \{0,1\}^n$. Recall that the volume of the Hamming ball with radius $r$ is given by

$$V(n,r) = \sum_{i=0}^{r} \binom{n}{i}. \tag{0.44}$$

The metric entropy for $\{0,1\}^n$ at integer resolution $\delta$, $0 < \delta < n/2$, can be bounded using the same arguments as in (0.43) to obtain

$$\log\Big(\frac{2^n}{V(n,\delta)}\Big) \le H\big(\delta; \{0,1\}^n, \rho_{\mathrm{H}}\big) \le \log\Big(\frac{2^n}{V(n,\lfloor\delta/2\rfloor)}\Big). \tag{0.45}$$

Note the correspondence of the lower and upper bounds, respectively, to the Hamming bound and the Gilbert–Varshamov bound in coding theory. We can also establish the following upper bound

$$H\big(\delta; \{0,1\}^n, \rho_{\mathrm{H}}\big) \le \log\Big(\frac{2^n(n\ln 2 + 1)}{V(n,\delta)}\Big), \tag{0.46}$$

by showing the existence of a $\delta$-net of size at most $(2^n/V(n,\delta))(n\ln 2+1)$. To do so, we randomly and independently select $k$ points according to the uniform pmf over $\{0,1\}^n$. Then the probability that any fixed point in $\{0,1\}^n$ is not contained in a Hamming ball of radius $\delta$ centered at one of the chosen points is

$$\Big(1 - \frac{V(n,\delta)}{2^n}\Big)^k \le \exp\Big(-\frac{kV(n,\delta)}{2^n}\Big).$$

By the union bound,

$$\mathsf{Pr}\{\text{some point in } \{0,1\}^n \text{ is uncovered}\} \le 2^n \exp\Big(-\frac{kV(n,\delta)}{2^n}\Big).$$

Choosing $k = \lceil (2^n/V(n,\delta))\ln 2^n \rceil$ ensures that the probability above is strictly less than 1. Consequently, such a $\delta$-net exists, and the metric entropy is bounded as stated in (0.46).

**Relationship to the rate–distortion function.** The concept of covering introduced above can be viewed as a deterministic analogue of the randomized covering in the coding scheme of the lossy compression theorem in. In this analogy, the covering number corresponds to the smallest number of reproduction sequences required to achieve a given distortion level. The limiting per-dimension metric entropy, as the number of dimensions grows, plays the role of the rate-distortion function. In lossy compression, we allow a set of source sequences with vanishing probability (nontypical sequences) to remain uncovered, whereas in the metric entropy setup the covering is required to be exact. Consequently, for a given appropriately defined source and distortion measure, the limiting per-dimension metric entropy is expected to be an upper bound on the rate–distortion function. To make this argument concrete, consider the above two examples.

We first compare the per-dimension metric entropy of the Euclidean ball to the quadratic Gaussian rate–distortion function $R(D) = (1/2)\log(P/D)$, $0 \le D \le P$. Setting $r = \sqrt{nP}$ and $\delta = \sqrt{nD}$ in (0.43) and dividing by $n$, we obtain

$$\frac{1}{2}\log\left(\frac{P}{D}\right) \le \frac{1}{n}H\left(\sqrt{nD}; B(\sqrt{nP}), \rho\right) \le \log\left(1 + 2\sqrt{\frac{P}{D}}\right) \tag{0.47}$$

and the per-dimension metric entropy is lower bounded by $R(D)$ for every $n > 1$.

Now we compare the limiting per-dimension metric entropy of the Hamming cube to the rate distortion function of the binary symmetric source with Hamming distortion given by $R(D) = 1 - \mathsf{H}(D)$, $0 \le D \le 1/2$. Setting $\delta = nD$ and using Stirling's approximation, it follows that

$$\frac{1}{\sqrt{8nD(1-D)}}2^{n\,\mathsf{H}(D)} \le V(n, nD) \le 2^{n\,\mathsf{H}(D)}. \tag{0.48}$$

Substituting in the lower bound in (0.45) and the upper bound in (0.46) and dividing by $n$, we obtain

$$1 - \mathsf{H}(D) \le \frac{1}{n}H\left(nD; \{0,1\}^n, \rho_{\mathrm{H}}\right) \le 1 - \mathsf{H}(D) + O(\log n/n). \tag{0.49}$$

Taking the limit as $n \to \infty$, we have shown that the limiting per-dimension metric entropy of the Hamming cube is equal to the rate–distortion function of the binary symmetric source with Hamming distortion.

Examples 0.3 and 0.4 are of finite-dimensional metric spaces. As the following example shows, metric entropy can be finite for arbitrary metric spaces.

**Example 0.5** (Lipschitz functions on unit interval)**.** Consider the set of functions $\mathscr{F} = \{\phi : [0,1] \to \mathbb{R},\ \phi(0) = 0,\ |\phi(v) - \phi(v')| \le L|v - v'| \text{ for all } v, v' \in [0,1]\}$, for some fixed constant $L > 0$, with the metric $\rho_\infty(\phi, \phi') = \sup_{v\in[0,1]} |\phi(v) - \phi'(v)|$. The metric entropy for $\mathscr{F}$ at resolution $\delta > 0$ satisfies

$$\left\lfloor \frac{L}{\delta} \right\rfloor \le H(\delta; \mathscr{F}, \rho_\infty) \le \left\lceil \frac{2L}{\delta} \right\rceil \log 3. \tag{0.50}$$

To prove the lower bound, assume that $0 < \delta < L$ and partition the interval $[0,1]$ into $m = \lfloor L/\delta \rfloor$ equal-sized subintervals of length $\Delta = 1/m$. For each $\beta = (\beta_1, \beta_2, \dots, \beta_m) \in \{-1,1\}^m$, define $\phi_\beta$ to be a piecewise linear function with $\phi_\beta(0) = 0$ and $d\phi_\beta(v)/dv = \beta_j L$ on subinterval $j = 1, 2, \dots, m$ (see Figure 0.1).

For $\beta \neq \beta'$, let $j$ be the smallest index where they differ. Then on the $j$-th subinterval one function goes up by $L\Delta$ and the other goes down by $L\Delta$. Hence,

$$\sup_v |\phi_\beta(v) - \phi_{\beta'}(v)| \geq 2L\Delta \tag{0.51}$$

$$\geq \frac{2L}{m} \tag{0.52}$$

$$\geq \frac{2L}{L/\delta} \tag{0.53}$$

$$\geq 2\delta. \tag{0.54}$$

Thus, any $\delta$-net of $\mathscr{F}$ can contain at most one member of $\{\phi_\beta\colon \beta \in \{-1,1\}^m\}$ in each ball of radius $\delta$, and we obtain the lower bound

$$N(\delta; \mathscr{F}, \rho_\infty) \geq 2^m, \tag{0.55}$$

and the metric entropy is lower bounded as $H(\delta; \mathscr{F}, \rho_\infty) \geq \lfloor L/\delta \rfloor$. The upper bound can be proved using a similar approach and is left as an exercise.

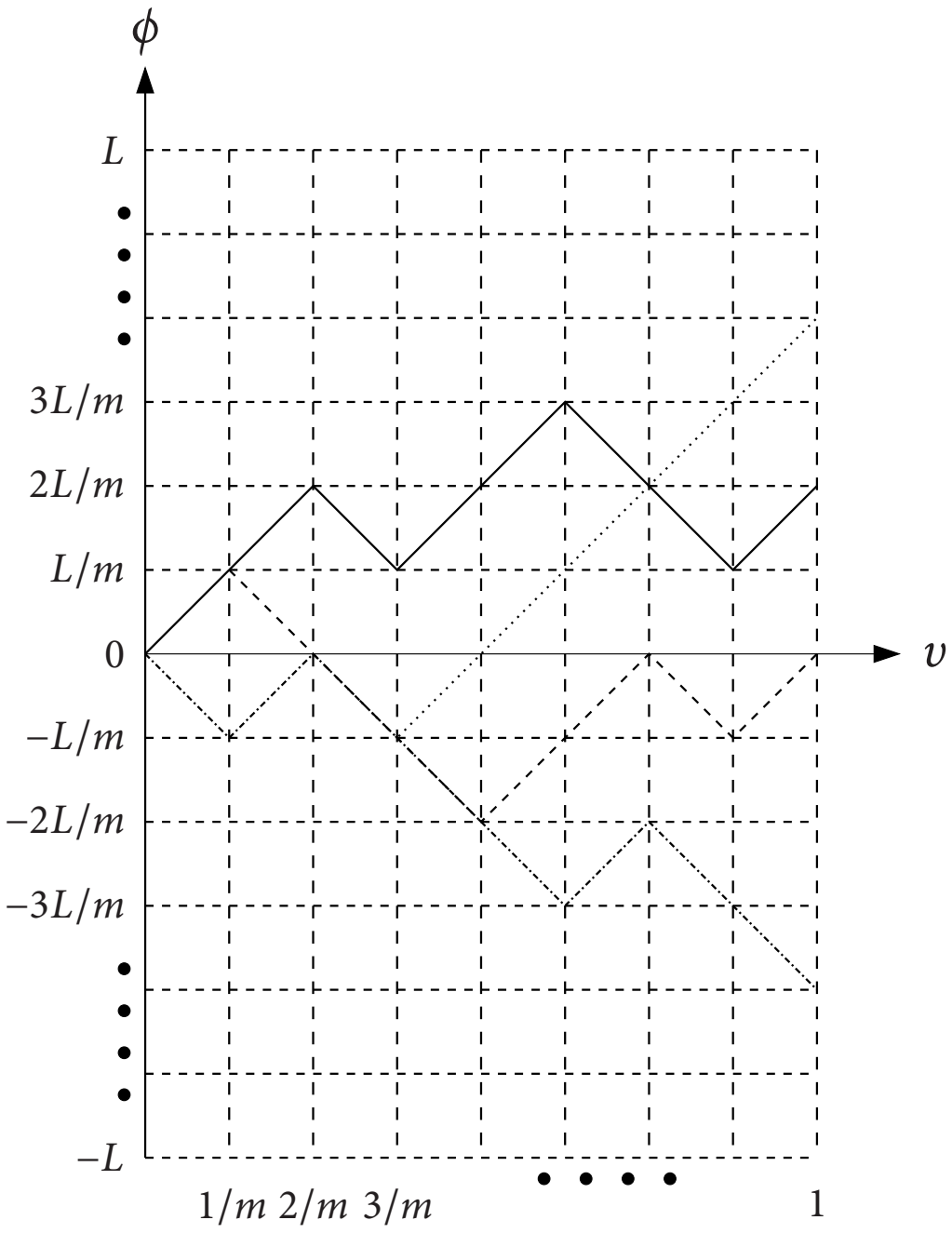

**Figure 0.1.** Piecewise Lipschitz functions.

**Remarks.**

1. Unlike in finite-dimensional examples, where metric entropy scales as $\log(1/\delta)$, in the infinite-dimensional example it grows much faster, at a rate of $1/\delta$; that is, the number of spheres needed to cover the space increases exponentially rather than polynomially as $\delta$ decreases.

2. We can readily extend the definition of covering numbers and metric entropy to a space with a *pseudometric* (also called *semimetric*) that satisfies all the requirements of a metric, except that it may have $\rho(x, y) = 0$ for some $x \neq y$.

We can similarly define the notion of *packing* as a dual to covering, and as the deterministic analogue of the random packing used to establish the channel coding theorem. In fact, we have already made use of it in Examples 0.3 and 0.4.

**Definition 0.6.** Given a metric space $(\mathscr{V}, \rho)$, a $\delta$-*packing* of a set $\mathscr{S} \subseteq \mathscr{V}$ is a set of points $\{v_1, v_2, \ldots, v_m\} \subseteq \mathscr{S}$ such that $\rho(v_i, v_j) > \delta$ for every $i \neq j$. The *packing number* of the set $\mathscr{S}$ at resolution $\delta$, $M(\delta; \mathscr{S}, \rho)$, is the maximum $m$ such that there exists a $\delta$-packing with $m$ points.

We can establish the following basic relationship between the covering and packing numbers of a set.

**Lemma 0.6.** For every $\delta > 0$, the covering and packing numbers of a set $\mathscr{S}$ at resolution $d$ are related as

$$M(2\delta; \mathscr{S}, \rho) \leq N(\delta; \mathscr{S}, \rho) \leq M(\delta; \mathscr{S}, \rho).$$

The proof of this lemma is left as an exercise.

## 0.3 LEARNING-THEORETIC SETUP

We introduce the learning-theoretic framework, which underlies the bounds derived in the following three sections. In the previous chapter we formulated the statistical learning problem as follows: Given a dataset consisting of i.i.d. samples from a fixed but unknown probability distribution $p_{\text{data}}$, select a probability distribution $p_{\hat{\theta}}$ from a model $\mathscr{P} = \{p_\theta : \theta \in \Theta\}$ that is closest to $p_{\text{data}}$. We saw that determining the model parameters that best fit the dataset reduces to optimizing a deterministic objective function. For example, in linear regression this corresponds to minimizing a quadratic objective function, while in logistic regression we minimize the log loss function. In learning theory, we adopt a more abstract formulation: the goal is to minimize excess risk with respect to a general loss function, which need not arise from a probability distance measure between the model and the data distribution.

### 0.3.1 Definitions and examples

**Definition 0.7.** Given a *data space* $\mathscr{Z}$, a *model class* (also referred to as *hypothesis class*) $\mathscr{W}$, a *loss function* $\ell : \mathscr{Z} \times \mathscr{W} \to \mathbb{R}_+$, and a probability distribution $p$ on the data space $\mathscr{Z}$, we define the *expected loss* (also called *risk*) of the model $w \in \mathscr{W}$ as $L(w) = \mathsf{E}_Z\left(\ell(Z, w)\right)$, where the expectation is with respect to $Z \sim p$. The minimum expected loss is defined as $L^*(\mathscr{W}) = \inf_{w \in \mathscr{W}} L(w)$.

For simplicity, we assume that the minimum exists and is achieved by some $w^* \in \mathscr{W}$, hence we have $L^*(\mathscr{W}) = L(w^*)$.

This learning framework covers many statistical learning problems. Consider the following examples.

**Binary classification.** Here $\mathscr{Z} = \mathscr{X} \times \mathscr{Y}$, where $\mathscr{X}$ is an arbitrary feature space and $\mathscr{Y} = \{0, 1\}$; $\mathscr{W}$ is a given class of functions $w : \mathscr{X} \to \{0, 1\}$; and $\ell$ is the $0 - 1$ loss function

$$\ell(z, w) = \ell((x, y), w) = \mathbb{1}_{\{y \neq w(x)\}}.$$

The elements of $\mathscr{W}$ are typically referred to as (binary) *classifiers*. The risk of $w \in \mathscr{W}$ is the probability of error

$$L(w) = \mathsf{Pr}\{Y \neq w(X)\}.$$

For the special case of a *linear classifier* (or perceptron), $\mathscr{X} = \mathbb{R}^K$ and $w(\mathbf{x}) = \mathbb{1}_{\{\mathbf{w}^T\mathbf{x} \geq 0\}}$, where $\mathbf{w}^T = [w_0\ w_1\ \cdots\ w_K]$ and $\mathbf{x}^T = [1\ x_1\ \cdots\ x_K]$.

**Regression.** Here $\mathscr{Z} = \mathscr{X} \times \mathscr{Y}$, where $\mathscr{X}$ is an arbitrary feature space and $\mathscr{Y} \subseteq \mathbb{R}$; $\mathscr{W}$ is a given class of functions $w : \mathscr{X} \to \mathbb{R}$; and $\ell$ is the squared error loss

$$\ell(z, w) = \ell((x, y), w) = (y - w(x))^2.$$

The risk of $w \in \mathscr{W}$ is its mean squared error (MSE),

$$L(w) = \mathsf{E}_{X,Y}\big((Y - w(X))^2\big).$$

For the special case of linear regression discussed in Section **??**, $\mathscr{X} = \mathbb{R}^K$ and $w(\mathbf{x}) = \mathbf{w}^T\mathbf{x}$, where $\mathbf{w}$ and $\mathbf{x}$ are the same as defined above. The risk is $L(w) = \mathsf{E}\big((Y - \mathbf{w}^T\mathbf{X})^2\big)$.

**Logistic regression.** Here $\mathscr{Z} = \mathscr{X} \times \mathscr{Y}$, where $\mathscr{X} = \mathbb{R}^K$ and $\mathscr{Y} = \{0, 1\}$, $\mathscr{W}$ is the class of functions

$$w(\mathbf{x}) = \sigma\Big(w_0 + \sum_{j=1}^{K} w_j x_j\Big), \tag{0.56}$$

where $\sigma$ is the logistic function defined in Section **??**, $\ell$ is the log-loss function

$$\ell((\mathbf{x}, y), w) = -y \log w(\mathbf{x}) - (1 - y)\log(1 - w(\mathbf{x})),$$

and $w \in [0, 1]$ represents the probability that that $Y = 1$. Defining $\mathsf{E}(Y|\mathbf{X} = \mathbf{x}) = p(\mathbf{x})$, the risk for $w \in \mathscr{W}$ is

$$L(w) = \mathsf{E}\big(-p(\mathbf{X})\log w(\mathbf{X}) - (1 - p(\mathbf{X}))\log(1 - w(\mathbf{X}))\big),$$

which is the expected cross-entropy between $(p(\mathbf{X}), 1 - p(\mathbf{X}))$ and $(w(\mathbf{X}), 1 - w(\mathbf{X}))$.

**Clustering (vector quantization).** Here $\mathscr{Z} = \mathbb{R}^d$, $\mathscr{W}$ is the set of all $k$-element subsets of $\mathcal{X}$, and $\ell$ is the nearest-neighbor squared error given by $\ell(z, w) = \min_{a \in w} \|a - z\|_2^2$. The risk of $w \in \mathcal{W}$ is the expected distortion

$$L(w) = \mathsf{E}\Big(\min_{a \in w} \|Z - a\|_2^2\Big). \tag{0.57}$$

In the above examples, the risk $L(w)$ of each $w$ is computed with respect to a fixed probability distribution $p$ on $\mathcal{X}$, which plays the role of $p_{\text{data}}$. The learning problem arises when the data distribution is not known and we use a dataset to fit the model. To formulate this problem precisely, we need the following definitions.

**Definition 0.8.** Let $Z_1, Z_2, \ldots, Z_n$ be an i.i.d. dataset drawn from $p$. We define the *empirical risk* for a model $w \in \mathcal{W}$ as

$$L_n(w) = \frac{1}{n}\sum_{i=1}^{n} \ell(Z_i, w). \tag{0.58}$$

Since $\mathsf{E}_{Z^n}(L_n(w)) = L(w)$, the empirical loss is an unbiased estimate of $L(w)$ for every $w \in \mathcal{W}$.

**Definition 0.9.** A *learning algorithm* is a random mapping (channel) $Z^n \to W$ that takes the dataset $Z^n$ as input and produces a random variable $W$ taking values in $\mathcal{W}$.

**Definition 0.10.** Let $Z^n \to W$ be a learning algorithm and let the sample $Z \sim p_{\text{data}}$ be independent of the dataset $Z^n$, hence also independent of $W$. Define the *risk of* $W$ as $L(W) = \mathsf{E}_Z(\ell(Z, W))$. The difference between the risk of $W$ and its empirical risk, namely $L(W) - L_n(W)$, is referred to as the *generalization error*.

## 0.3.2 Learning problem

Given an i.i.d. dataset $Z^n$, the learning problem is to find a model $W$ taking values in $\mathcal{W}$, such that the *excess risk* $L(W) - L(w^*)$ is small either in expectation or with high probability. This is known as the *PAC learning framework* (Valiant 1984, Haussler 1992), where PAC stands for "Probably Approximately Correct." The name reflects the goal of designing learning algorithms that are *approximately correct*, that is, have small excess risk $L(W) - L(w^*)$ with high probability.

It is not difficult to see that the generalization error is closely related to the objective of making $L(W)$ close to $L(w^*)$ by expressing the excess risk as

$$L(W) - L(w^*) = \big(L(W) - L_n(W)\big) + \big(L_n(W) - L_n(w^*)\big) + \big(L_n(w^*) - L(w^*)\big). \tag{0.59}$$

Observe that:

1. The first term $(L(W) - L_n(W))$ is the generalization error of $W$.
2. The second term $(L_n(W) - L_n(w^*))$ is the difference between the empirical loss of $W$ and the empirical loss of the unknown optimal $w^*$ on the same dataset. It quantifies the ability of the learning algorithm to approximate $w^*$ based on the dataset $Z^n$.

3. The third term $(L_n(w^*) - L(w^*))$ is the difference between the empirical loss and its expected value for the minimizer $w^*$. This term, which is independent of the learning algorithm, measures the fluctuations of the empirical loss $L_n(w^*)$ around its expected value $L(w^*)$.

In statistical learning theory, we typically seek bounds on the excess risk that are independent of the unknown data distribution. Nevertheless, as we will see, some mild assumptions on the data-generating distribution and the loss function are still required.

To that end, we first use the tools developed in Section 0.1 to understand the behavior of the third term in (0.59). Suppose that the data distribution of $Z$ is such that, for every $w \in \mathscr{W}$, the random variables $\ell(Z, w)$ are $\sigma^2$-subgaussian. In particular, this holds for $w = w^*$. Under this assumption we obtain the following convergence result.

**Lemma 0.7.** $|L_n(w^*) - L(w^*)| \to 0$ with probability 1.

**Proof.** Since $\ell(Z_i, w^*)$ is $\sigma^2$-subgaussian, the scaled variables $\frac{1}{n}\ell(Z_i, w^*)$ are $\sigma^2/n^2$-subgaussian. Therefore, their sum $L_n(w^*)$ is $\sigma^2/n$-subgaussian (see Problem 0.2). Therefore, for any $\epsilon > 0$,

$$\mathsf{Pr}\left\{|L_n(w^*) - L(w^*)| \geq \epsilon\right\} \leq 2e^{-n\epsilon^2/2\sigma^2}. \tag{0.60}$$

Now let $Z_1, Z_2, \ldots$ be an i.i.d. sequence of samples from $p$. Then using the Borel–Cantelli lemma, it follows that for every $\epsilon > 0$,

$$\mathsf{Pr}\left\{|L_n(w^*) - L(w^*)| \geq \epsilon \text{ for infinitely many } n\right\} = 0. \tag{0.61}$$

This, in turn, implies that $L_n(w^*)$ converges to $L(w^*)$ almost surely, and the bound of (0.60) shows that this convergence is exponentially fast. This completes the proof.

### 0.3.3 Generalization error and ERM

We now take a closer look at the first two terms in (0.59). For the first term, we denote the *worst-case generalization error* by

$$\Delta_n(\mathscr{W}) = \sup_{w \in \mathscr{W}} |L(w) - L_n(w)|. \tag{0.62}$$

We will develop techniques for deriving upper bounds on the expected value of this quantity,

$$\mathsf{E}(\Delta_n(\mathscr{W})) = \mathsf{E}\big(\sup_{w \in \mathscr{W}} |L(w) - L_n(w)|\big), \tag{0.63}$$

where the expectation is taken with respect to $Z, Z^n$.

This quantity provides and upper bound the expected generalization error $\mathsf{E}(L(W) - L_n(W))$ for *any* learning algorithm. To see this, observe that for any $p(w|z^n)$ we have

$$\mathsf{E}_{W,Z^n,Z}\Big(\ell(Z, W) - \frac{1}{n}\sum_{i=1}^n \ell(Z_i, W)\Big) = \mathsf{E}_{Z^n,Z}\Big[\,\mathsf{E}\Big(\ell(Z, W) - \frac{1}{n}\sum_{i=1}^n \ell(Z_i, W)\Big|Z^n, Z\Big)\Big]$$

$$\leq \mathsf{E}_{Z^n,Z}\left[\max_{w\in\mathcal{W}}\left(\ell(Z,w)-\frac{1}{n}\sum_{i=1}^{n}\ell(Z_i,w)\right)\right].$$

Although bounding $\mathsf{E}(\Delta_n(\mathcal{W}))$ may appear loose for any specific learning algorithm, it is in fact well-suited for the following important learning algorithm.

**Definition 0.11.** The *Empirical Risk Minimization (ERM)* learning algorithm is defined as

$$w_{\mathrm{ERM}} = \arg\min_{w\in\mathcal{W}} L_n(w) = \arg\min_{w\in\mathcal{W}} \frac{1}{n}\sum_{i=1}^{n}\ell(Z_i,w). \tag{0.64}$$

**Lemma 0.8.** For the ERM,

$$\mathsf{E}\left(L(w_{\mathrm{ERM}})-L(w^*)\right) \leq \mathsf{E}(\Delta_n(\mathcal{W})), \tag{0.65}$$

where the expectation on the left-hand side is over two sources of randomness: the training data $Z^n$ and an independent sample $Z$. This can be written explicitly as

$$\mathsf{E}\left(L(w_{\mathrm{ERM}})-L(w^*)\right) = \mathsf{E}_{Z^n}\left[\,\mathsf{E}_Z\left(\ell(Z,w_{\mathrm{ERM}})-\ell(Z,w^*)\middle|Z^n\right)\right] \tag{0.66}$$

with $Z$ independent of $Z^n$ and drawn from the same distribution.

**Proof.** Since the inequality $L_n(w_{\mathrm{ERM}}) \leq L_n(w^*)$ holds for ERM, the second term in (0.59) is less than or equal to zero. Consequently, the expected excess risk of ERM can be upper-bounded as

$$\begin{aligned}
\mathsf{E}\left(L(w_{\mathrm{ERM}})-L(w^*)\right) &\leq \mathsf{E}\left(L(w_{\mathrm{ERM}})-L_n(w_{\mathrm{ERM}})\right)+\mathsf{E}\left(L_n(w^*)-L(w^*)\right)\\
&= \mathsf{E}\left(L(w_{\mathrm{ERM}})-L_n(w_{\mathrm{ERM}})\right) && (0.67)\\
&\leq \mathsf{E}(\Delta_n(\mathcal{W})). && (0.68)
\end{aligned}$$

This completes the proof.

In Section 0.4, we use concentration inequalities with metric entropy bounds to show that, in a wide variety of settings, the expected value of $\Delta_n(\mathcal{W})$ satisfies

$$\mathsf{E}(\Delta_n(\mathcal{W})) \leq \mathrm{const}\cdot\sqrt{\frac{C(\mathcal{W})}{n}}, \tag{0.69}$$

where the quantity $C(\mathcal{W})$ depends only on the loss function and the model class, capturing the "learning capacity" of $\mathcal{W}$. Consequently, whenever $C(\mathcal{W})$ is well-defined and finite, the ERM scheme is *consistent* in the sense that its excess risk vanishes at the rate $O(1/\sqrt{n})$ as $n\to\infty$.

Observe that, as part of proving Lemma 0.8, we have also shown that, for $W = w_{\mathrm{ERM}}$, the second term in (0.59) is less than or equal to zero. For other algorithms $W$, the second term in (0.59) is the optimality gap of $W$ compared to $w_{\mathrm{ERM}}$ on a given sample:

$$L_n(W)-L_n(w^*) \leq L_n(W) - \inf_{w\in\mathcal{W}} L_n(w)$$

$$= L_n(W) - L_n(w_{\text{ERM}}). \tag{0.70}$$

This optimality gap must be analyzed on a case-by-case basis, for each specific algorithm. Further details can be found in (Bottou and Bousquet 2007).

The use of ERM as a benchmark in linear and logistic regression is justified because, in these settings, the empirical risk can be minimized exactly. In more complex model classes—such as neural networks— training is typically carried out using stochastic gradient descent, as discussed in the previous chapter. In this case, the mapping from $Z^n$ to $W$ is *random*, due to the randomness introduced by sampling in each gradient update, and the empirical risk need not be minimized. Although the primary quantity of interest remains the excess risk, controlling the generalization error helps ensure that the learned model's predictions on new data reflect the underlying (unobservable) population distribution, rather than spurious patterns peculiar to the training sample.

In Section 0.5, we use information theory to develop a bound on the expected generalization error $\mathsf{E}(L(W) - L_n(W))$ for such randomized learning algorithms. Instead of the "worst-case" capacity $C(\mathscr{W})$, we will make use of the mutual information $I(Z^n; W)$ between the training dataset $Z^n$ and the output $W$ of the learning algorithm.

In Section 0.6, we briefly examine an alternative framework for analyzing randomized learning algorithms: the *PAC-Bayes* framework due to McAllester (1999). Similar to the classical PAC framework, PAC-Bayes theory provides conditions under which a learning algorithm attains a small generalization error with high probability. It differs from standard PAC learning theory in that it uses information-theoretic measures to quantify and control the learning algorithm's capacity.

## 0.4 BOUNDS ON WORST-CASE GENERALIZATION ERROR

We have already shown that the expected excess risk of ERM can be upper-bounded as $\mathsf{E}\left(L(w_{\text{ERM}}) - L(w^*)\right) \le \mathsf{E}(\Delta_n(\mathscr{W}))$. We now turn to the analysis of the expected value of $\Delta_n(\mathscr{W})$.

Throughout this section, we will assume that $\ell(Z, w)$ is $\sigma^2$-subgaussian for every $w \in \mathscr{W}$ and some $\sigma > 0$. While this assumption involves the unknown probability distribution $p$ of $Z$, it can often be verified without knowing $p$, for example, when $\ell$ is bounded.

### 0.4.1 Generalization bound for finite model classes

We can use the maximal inequalities presented in Section 0.1 to bound $\mathsf{E}\left(\Delta_n(\mathscr{W})\right)$ as follows. For each $w \in \mathscr{W}$, $L_n(w)$ is a sum of i.i.d. random variables $\frac{1}{n}\ell(Z_i, w)$, each of which is $(\sigma^2/n^2)$-subgaussian for some $\sigma^2 > 0$. Thus, $L_n(w)$ is $(\sigma^2/n)$-subgaussian for every $w \in \mathscr{W}$. Using (0.36) and the fact that $\log(2|\mathscr{W}|) \le 2\log|\mathscr{W}|$ for $|\mathscr{W}| \ge 2$, we immediately obtain the following bound.

**Theorem 0.1.** Let $\mathscr{W}$ be a finite model class and $Z_1, Z_2, \ldots, Z_n$ be i.i.d. samples. Then,

$$\mathsf{E}\left(\Delta_n(\mathscr{W})\right) \le 2\sigma\sqrt{\frac{\log|\mathscr{W}|}{n}}. \tag{0.71}$$

There are two key insights here. The first is the $\sqrt{\log|\mathscr{W}|}$ dependence of the excess risk on the size of $\mathscr{W}$. We can think of the size of $\mathscr{W}$ as a measure of its "capacity," reflecting the intuition that larger model classes tend to be more expressive. The second insight is the $1/\sqrt{n}$ decay of the excess risk with the size of the training set—the more samples we collect, the closer the performance of the empirically optimal predictor is to the theoretical optimum.

### 0.4.2 Generalization bound for infinite model classes

While the generalization bound for finite model classes highlights some of the main ideas, most applications, including the examples in Section 0.3, involve infinite model classes. To make the analysis of infinite model classes more tractable, we first consider the special case in which a distance (or metric) is defined on $\mathscr{W}$, such that, if $w$ and $w'$ are close to each other, their corresponding empirical losses $L_n(w)$ and $L_n(w')$ are also close for every outcome $z^n$ of the dataset. We can then "quantize" $\mathscr{W}$ to some small resolution, i.e., replace it by a finite approximation $\hat{\mathscr{W}}$. This then leads to upper bounds on risk that scale roughly as $\sqrt{\log|\hat{\mathscr{W}}|/n}$.

Formally, we introduce a metric $\rho$ on the model class $\mathscr{W}$ such that the following two conditions are satisfied:

1. $(\mathscr{W}, \rho)$ is a totally bounded metric space.
2. There exists a constant $c > 0$ such that, for all $w, w' \in \mathscr{W}$ and for every dataset outcome $z^n$, the following Lipschitz condition is satisfied:

$$\left|\frac{1}{n}\sum_{i=1}^{n} \ell(z_i, w) - \frac{1}{n}\sum_{i=1}^{n} \ell(z_i, w')\right| \leq c\rho(w, w'). \tag{0.72}$$

Now, for $\delta > 0$ and $\delta' = \delta/c$, let $\hat{\mathscr{W}} = \{w_1, w_2, \ldots, w_N\}$ be a $\delta'$-net for $\mathscr{W}$ of minimal cardinality; that is, $N = N(\delta'; \mathscr{W}, \rho)$. Further, for each $w \in \mathscr{W}$, let $w(\delta)$ be its nearest neighbor in $\{w_1, w_2, \ldots, w_N\}$, i.e., $w(\delta) = w_i$ for some $i \in \{1, 2, \ldots, N\}$ and $\rho(w, w(\delta)) = \rho(w, w_i) = \min_{1\leq j\leq N} \rho(w, w_j) \leq \delta'$. Then

$$\begin{aligned}
\mathsf{E}\Big(\sup_{w\in\mathscr{W}} |L_n(w) - L(w)|\Big) &\leq \mathsf{E}\Big(\sup_{w\in\mathscr{W}} |L_n(w) - L_n(w(\delta))|\Big) \\
&\quad + \sup_{w\in\mathscr{W}} |L(w) - L(w(\delta))| + \mathsf{E}\Big(\sup_{w\in\mathscr{W}} |L_n(w(\delta)) - L(w(\delta))|\Big) &(0.73)\\
&= \mathsf{E}\Big(\sup_{w\in\mathscr{W}} |L_n(w) - L_n(w(\delta))|\Big) + \sup_{w\in\mathscr{W}} |L(w) - L(w(\delta))| + \mathsf{E}\big(\Delta_n(\hat{\mathscr{W}})\big), &(0.74)
\end{aligned}$$

where each of the three terms on the right-hand side can be bounded as follows. For the first term, we have

$$\begin{aligned}
\mathsf{E}\Big(\sup_{w\in\mathscr{W}} |L_n(w) - L_n(w(\delta))|\Big) &\leq c \sup_{w\in\mathscr{W}} \rho(w, w(\delta)) \\
&= c \cdot \sup_{w\in\mathscr{W}} \min_{1\leq i\leq N} \rho(w, w_i)
\end{aligned}$$

$$\le c \cdot (\delta/c)$$
$$= \delta.$$

For the second term, we have

$$\begin{aligned}
\sup_{w \in \mathcal{W}} |L(w) - L(w(\delta))| &= \sup_{w \in \mathcal{W}} |\,\mathsf{E}(L_n(w) - L_n(w(\delta)))| && (0.75)\\
&\le \sup_{w \in \mathcal{W}} \mathsf{E}\,|L_n(w) - L_n(w(\delta))| && (0.76)\\
&\le c \sup_{w \in \mathcal{W}} \rho(w, w(\delta)) && (0.77)\\
&\le \delta, && (0.78)
\end{aligned}$$

where (0.76) follows from Jensen's inequality. Finally, for the third term, from Theorem 0.1, it follows that

$$\mathsf{E}\left(\Delta_n(\hat{\mathcal{W}})\right) \le 2\sigma\sqrt{\frac{\log N(\delta'; \mathcal{W}, \rho)}{n}}. \tag{0.79}$$

Optimizing over $\delta$, we have established the following generalization bound.

**Lemma 0.9.** Let $(\mathcal{W}, \rho)$ be a totally bounded metric space and assume the Lipschitz continuity assumption in (0.72) holds. Then for $\delta > 0$, $\delta' = \delta/c$, and $n$ i.i.d. samples, we have

$$\mathsf{E}(\Delta_n(\mathcal{W})) \le \min_{\delta > 0}\left(2\delta + 2\sigma \cdot \sqrt{\frac{H(\delta'; \mathcal{W}, \rho)}{n}}\right). \tag{0.80}$$

The Lipschitz continuity condition in (0.72) is too restrictive since it must hold for every possible outcome of the dataset. However, the main idea underlying it—namely, that it should be possible to split $\mathcal{W}$ into finitely many subsets $\mathcal{W}_1, \mathcal{W}_2, \ldots, \mathcal{W}_N$ such that the empirical losses $L_n(w)$ with $w$ in each $\mathcal{W}_i$ are somehow "close" to each other—still provides the correct intuition. The challenge in applying it directly is that, in general, the partition of $\mathcal{W}$ into the sets $\mathcal{W}_1, \mathcal{W}_2, \ldots, \mathcal{W}_N$ will depend on the outcome of the dataset. However, as we will see next, the fact that we are dealing with averages of i.i.d. random variables provides us with a great deal of symmetry that can be exploited by introducing additional randomness.

### 0.4.3 Rademacher Complexity

We introduce the Rademacher complexity, a measure of the richness of a function class, and use it to derive a general bound on worst-case generalization error. Although this quantity can be defined for any function class, we present it here in terms of the loss function for a learning problem.

**Definition 0.12.** Consider a learning problem with data space $\mathcal{Z}$, model class $\mathcal{W}$, loss function $\ell(z, w)$, and a dataset $Z_1, Z_2, \ldots, Z_n$ drawn i.i.d. from a distribution $p$. Let $\varepsilon_1, \varepsilon_2, \ldots, \varepsilon_n$ be i.i.d. Rademacher random variables (as defined in Example 0.2) independent of $Z_1, Z_2, \ldots, Z_n$. Then the *empirical Rademacher complexity* for this learning problem is defined as

$$\bar{\mathsf{R}}_n(\mathscr{W}, Z^n) = \mathsf{E}_{\varepsilon^n}\left(\sup_{w\in\mathscr{W}}\left|\frac{1}{n}\sum_{i=1}^{n}\varepsilon_i \ell(Z_i, w)\right|\right), \tag{0.81}$$

and the *Rademacher complexity* is its expectation with respect to $Z^n$, that is,

$$\mathsf{R}_n(\mathscr{W}) = \mathsf{E}_{\varepsilon^n, Z^n}\left(\sup_{w\in\mathscr{W}}\left|\frac{1}{n}\sum_{i=1}^{n}\varepsilon_i \ell(Z_i, w)\right|\right). \tag{0.82}$$

In words, the Rademacher complexity is the expected maximum empirical correlation between the loss vector $(\ell(Z_1, w), \ldots, \ell(Z_n, w))$ and an independent "noise" vector $(\varepsilon_1, \ldots, \varepsilon_n)$, where the maximum is taken over all $w \in \mathscr{W}$.

Intuitively, this captures the idea that a model class is too large to support reliable learning if, with high probability, there exists a model whose losses align closely with random noise. Conversely, if the Rademacher complexity decays as the sample size $n$ grows, then no model in the class can, on average, correlate strongly with such noise, suggesting better generalization.

We illustrate the Rademacher complexity with the following examples.

**Example 0.6** (Finite $\mathscr{W}$)**.** Consider a learning problem with finite model class $\mathscr{W}$ and a loss function $\ell(Z, w)$ which is $\sigma^2$-subgaussian for each $w \in \mathscr{W}$. For a fixed dataset $Z^n$, the Rademacher complexity $\mathsf{R}_n(\mathscr{W})$ is the expectation of the maximum absolute value of $|\mathscr{W}|$ random variables

$$V(w) = \frac{1}{n}\sum_{i=1}^{n}\varepsilon_i \ell(Z_i, w). \tag{0.83}$$

Each $V(w)$ is, in turn, a sum of $n$ random variables $V_i(w) = \frac{1}{n}\varepsilon_i \ell(Z_i, w)$. Since $\varepsilon_i$ and $Z_i$ are independent, we have $\mathsf{E}(V_i(w)) = 0$ and, for any $\lambda > 0$

$$\begin{aligned}
\mathsf{E}\left(e^{\lambda V_i(w)}\right) &= \frac{1}{2}\mathsf{E}\left(e^{-(\lambda/n)\ell(Z_i,w)}\right) + \frac{1}{2}\mathsf{E}\left(e^{(\lambda/n)\ell(Z_i,w)}\right) && (0.84)\\
&\leq \exp\left(\frac{\lambda^2\sigma^2}{2n^2}\right). && (0.85)
\end{aligned}$$

Thus, $V(w)$ is a sum of $n$ i.i.d. $(\sigma/n)^2$-subgaussian random variables and is therefore $(\sigma^2/n)$-subgaussian. Then assuming $|\mathscr{W}| \geq 2$, from Problem 0.8, we have

$$\mathsf{R}_n(\mathscr{W}) = \mathsf{E}\left(\sup_{w\in\mathscr{W}}|V(w)|\right) \leq 2\sigma\sqrt{\frac{\log|\mathscr{W}|}{n}}. \tag{0.86}$$

Note that this is the same as the bound on worst-case generalization error in Theorem 0.1. We will make this relationship precise shortly.

For the next example we need the following important structural property of the Rademacher complexity due to Ledoux and Talagrand (1991).

**Theorem 0.2** (Contraction principle)**.** Let $\phi_1, \phi_2, \dots, \phi_n : \mathbb{R} \to \mathbb{R}$ be $n$ $c$-Lipschitz functions satisfying $\phi_i(0) = 0$ for each $i$. Let $\varepsilon_1, \varepsilon_2, \dots, \varepsilon_n$ be i.i.d. Rademacher random variables. Then, for any set $\mathscr{A} \subseteq \mathbb{R}^n$,

$$\mathsf{E}_{\varepsilon^n}\Big( \sup_{\mathbf{a}\in\mathscr{A}} \Big| \sum_{i=1}^n \varepsilon_i \phi_i(a_i) \Big| \Big) \le 2c\, \mathsf{E}_{\varepsilon^n}\Big( \sup_{\mathbf{a}\in\mathscr{A}} \Big| \sum_{i=1}^n \varepsilon_i a_i \Big| \Big). \tag{0.87}$$

**Example 0.7** (Linear regression)**.** Consider the linear regression setting in Section 0.3. Assume that $\mathscr{X}$ is a bounded subset of $\mathbb{R}^K$ and $\mathscr{Y}$ is a bounded interval in $\mathbb{R}$. The model class $\mathscr{W}$ consists of linear predictors of the form $w(\mathbf{x}) = \mathbf{w}^T\mathbf{x}$, where we assume that $\|\mathbf{w}\|_2 \le r$ for some $r > 0$. Then, for any i.i.d. dataset $Z_1 = (\mathbf{X}_1, Y_1), \dots, Z_n = (\mathbf{X}_n, Y_n)$ drawn from a probability distribution supported on $\mathscr{X} \times \mathscr{Y}$,

$$\bar{\mathsf{R}}_n(\mathscr{W}, Z^n) = \mathsf{E}_{\varepsilon^n}\Big( \frac{1}{n} \sup_{\mathbf{w}\in\mathbb{R}^K : \|\mathbf{w}\|\le r} \Big| \sum_{i=1}^n \varepsilon_i (Y_i - \mathbf{w}^T\mathbf{X}_i)^2 \Big| \Big). \tag{0.88}$$

Each term in the sum on the right-hand side of (0.88) contains the square of the prediction error $Y_i - \mathbf{w}^T\mathbf{X}_i$. Define the set $\mathscr{A} = \{(Y_1 - \mathbf{w}^T\mathbf{X}_1), \dots, (Y_n - \mathbf{w}^T\mathbf{X}_n) : \|\mathbf{w}\| \le r\} \subseteq \mathbb{R}^n$. Under the boundedness assumptions, the prediction errors are bounded by $b = \max_{y\in\mathscr{Y}} |y| + r \max_{\mathbf{x}\in\mathscr{X}} \|\mathbf{x}\|$. We now introduce the functions

$$\phi_1(u) = \ldots = \phi_n(u) = \begin{cases} u^2 & \text{for } |u| \le b, \\ b^2 & \text{otherwise.} \end{cases} \tag{0.89}$$

Note that these functions are $2b$-Lipschitz, since on the interval $[-b, b]$, the derivative of the function $\phi : u \to u^2$ is bounded by $2b$. Applying Theorem 0.2 gives

$$\bar{\mathsf{R}}_n(\mathscr{W}, Z^n) \le \frac{4b}{n} \mathsf{E}_{\varepsilon^n}\Big( \sup_{\mathbf{w}\in\mathbb{R}^K : \|\mathbf{w}\|\le r} \Big| \sum_{i=1}^n \varepsilon_i (Y_i - \mathbf{w}^T\mathbf{X}_i) \Big| \Big) \tag{0.90}$$

$$\le \frac{4b}{n}\Big( \mathsf{E}_{\varepsilon^n}\Big( \Big| \sum_{i=1}^n \varepsilon_i Y_i \Big| \Big) + \mathsf{E}_{\varepsilon^n}\Big( \sup_{\mathbf{w}\in\mathbb{R}^K : \|\mathbf{w}\|\le r} \Big| \sum_{i=1}^n \varepsilon_i \mathbf{w}^T\mathbf{X}_i \Big| \Big) \Big). \tag{0.91}$$

The first term does not involve a supremum, and can be bounded as follows:

$$\frac{1}{n} \mathsf{E}_{\varepsilon^n}\Big( \Big| \sum_{i=1}^n \varepsilon_i Y_i \Big| \Big) = \frac{1}{n} \mathsf{E}_{\varepsilon^n}\left( \sqrt{\Big( \sum_{i=1}^n \varepsilon_i Y_i \Big)^2} \right) \tag{0.92}$$

$$\le \frac{1}{n} \sqrt{\mathsf{E}_{\varepsilon^n}\Big( \sum_{i=1}^n \varepsilon_i Y_i \Big)^2} \tag{0.93}$$

$$= \frac{1}{n} \sqrt{\sum_{i=1}^n Y_i^2}, \tag{0.94}$$

where (0.93) follows from Jensen's inequality and the last step follows since $\mathsf{E}(\varepsilon_i \varepsilon_j) = \delta_{ij}$, where $\delta_{ij}$ is the Kronecker delta function. For the second term in (0.91), consider

$$
\begin{aligned}
\frac{1}{n}\,\mathsf{E}_{\varepsilon^n}\Big( \sup_{\mathbf{w}\in\mathbb{R}^K:\|\mathbf{w}\|\le r} \Big|\sum_{i=1}^{n} \varepsilon_i \mathbf{w}^T \mathbf{X}_i\Big|\Big) &= \frac{1}{n}\,\mathsf{E}_{\varepsilon^n}\Big( \sup_{\mathbf{w}\in\mathbb{R}^K:\|\mathbf{w}\|_2\le r} \Big|\mathbf{w}^T\Big(\sum_{i=1}^{n} \varepsilon_i \mathbf{X}_i\Big)\Big|\Big) && (0.95)\\
&\le \frac{1}{n}\,\mathsf{E}_{\varepsilon^n}\Big( \sup_{\mathbf{w}\in\mathbb{R}^K:\|\mathbf{w}\|_2\le r} \|\mathbf{w}\|_2 \Big\|\sum_{i=1}^{n} \varepsilon_i \mathbf{X}_i\Big\|_2\Big) && (0.96)\\
&= \frac{r}{n}\,\mathsf{E}_{\varepsilon^n}\Big( \Big\|\sum_{i=1}^{n} \varepsilon_i \mathbf{X}_i\Big\|_2\Big) && (0.97)\\
&= \frac{r}{n}\,\mathsf{E}_{\varepsilon^n}\left( \sqrt{\sum_{i,j=1}^{n} \varepsilon_i \varepsilon_j \mathbf{X}_i^T \mathbf{X}_j}\right) && (0.98)\\
&\le \frac{r}{n}\sqrt{\sum_{i=1}^{n} \|\mathbf{X}_i\|_2^2}\,, && (0.99)
\end{aligned}
$$

where (0.96) follows from the Cauchy–Schwarz inequality for vectors and the last step follows from Jensen's inequality and the fact that $\mathsf{E}(\varepsilon_i \varepsilon_j) = \delta_{ij}$. Combining the above bounds, we arrive at the following upper bound on the empirical Rademacher complexity for linear regression,

$$
\bar{\mathsf{R}}_n(\mathscr{W}, Z^n) \le \frac{4b}{n}\left( \sqrt{\sum_{i=1}^{n} Y_i^2} + r\sqrt{\sum_{i=1}^{n} \|\mathbf{X}_i\|_2^2}\right). \qquad (0.100)
$$

Since the training samples $(\mathbf{X}_i, Y_i)$ take values in a bounded set, the expected Rademacher complexity, hence the bound on the worst case generalization error, scales as $O(1/\sqrt{n})$. However, the bound (0.100) for the empirical Rademacher complexity says a bit more since it grows with the magnitude of the training data.

We now show that the expected generalization error of the learning problem can be upper bounded in terms of the Rademacher complexity.

**Theorem 0.3.** Let $\mathscr{W}$ be a model class, and let $Z_1, Z_2, \ldots, Z_n$ be i.i.d. samples. Then,

$$
\mathsf{E}(\Delta_n(\mathscr{W})) \le 2\,\mathsf{R}_n(\mathscr{W}). \qquad (0.101)
$$

**Proof.** We will introduce auxiliary random variables (as in Lemma 0.3) and use *symmetrization technique* introduced in the foundational work of Vapnik and Chervonenkis, though it has earlier roots in probability theory.

Let $\bar{Z}_1, \bar{Z}_2, \ldots, \bar{Z}_n$ be i.i.d. auxiliary samples from $p$ that are independent of $Z_1, Z_2, \ldots, Z_n$. Let $\bar{L}_n(w)$ be the empirical loss of $w$ on the ghost samples, i.e.,

$$
\bar{L}_n(w) = \frac{1}{n}\sum_{i=1}^{n} \ell(\bar{Z}_i, w). \qquad (0.102)
$$

Since $Z_i$ and $\bar{Z}_i$ have the same distribution, it follows that $L(w) = \mathsf{E}\left(\bar{L}_n(w)\right)$. Hence

$$\begin{aligned}
\mathsf{E}_{Z^n}(\Delta_n(\mathscr{W})) &= \mathsf{E}_{Z^n}\Big(\sup_{w\in\mathscr{W}} |L_n(w) - \mathsf{E}_{\bar{Z}^n}(\bar{L}_n(w))|\Big) && (0.103)\\
&= \mathsf{E}_{Z^n}\Big(\sup_{w\in\mathscr{W}} |\,\mathsf{E}_{\bar{Z}^n}(L_n(w) - \bar{L}_n(w))|\Big) && (0.104)\\
&\le \mathsf{E}_{Z^n}\Big(\sup_{w\in\mathscr{W}} \mathsf{E}_{\bar{Z}^n}\left(|(L_n(w) - \bar{L}_n(w))|\right)\Big) && (0.105)\\
&\le \mathsf{E}_{Z^n,\bar{Z}^n}\Big(\sup_{w\in\mathscr{W}} |L_n(w) - \bar{L}_n(w)|\Big), && (0.106)
\end{aligned}$$

where (0.105) follows from Jensen's inequality and the convexity of $|x|$. This device allows us to take advantage of the symmetry between the original samples and the ghost samples. Notice that, for each $w \in \mathscr{W}$, the random variables $U_i(w) = \ell(Z_i, w) - \ell(\bar{Z}_i, w)$, $i = 1, 2, \ldots, n$, are not only i.i.d. but also *symmetric*, i.e., $U_i(w)$ and $-U_i(w)$ have the same probability distribution. Consequently, if we introduce i.i.d. Rademacher random variables $\varepsilon_1, \varepsilon_2, \ldots, \varepsilon_n$ that are also independent of the $Z_i$'s and $\bar{Z}_i$'s, then the collection of random vectors $\left(\varepsilon_1 U_1(w), \varepsilon_2 U_2(w), \ldots, \varepsilon_n U_n(w)\right)$, $w \in \mathscr{W}$, has the same joint probability distribution as $\left(U_1(w), U_2(w), \ldots, U_n(w)\right)$, $w \in \mathscr{W}$.

The purpose of these constructions will now become clear. Consider

$$\begin{aligned}
\mathsf{E}\Big(\sup_{w\in\mathscr{W}} |L_n(w) - \bar{L}_n(w)|\Big) &= \mathsf{E}\left(\frac{1}{n}\sup_{w\in\mathscr{W}}\left|\sum_{i=1}^n U_i(w)\right|\right) && (0.107)\\
&= \mathsf{E}\left(\frac{1}{n}\sup_{w\in\mathscr{W}}\left|\sum_{i=1}^n \varepsilon_i U_i(w)\right|\right) && (0.108)\\
&= \mathsf{E}\left(\frac{1}{n}\sup_{w\in\mathscr{W}}\left|\sum_{i=1}^n \varepsilon_i(\ell(Z_i, w) - \ell(\bar{Z}_i, w))\right|\right) && (0.109)\\
&\le \mathsf{E}\left(\frac{1}{n}\sup_{w\in\mathscr{W}}\left|\sum_{i=1}^n \varepsilon_i \ell(Z_i, w)\right|\right)\\
&\qquad + \mathsf{E}\left(\frac{1}{n}\sup_{w\in\mathscr{W}}\left|\sum_{i=1}^n \varepsilon_i \ell(\bar{Z}_i, w)\right|\right) && (0.110)\\
&= 2\,\mathsf{E}\left(\frac{1}{n}\sup_{w\in\mathscr{W}}\left|\sum_{i=1}^n \varepsilon_i \ell(Z_i, w)\right|\right), && (0.111)
\end{aligned}$$

where the expectation in (0.108) is with respect to all the $Z_i$'s, $\bar{Z}_i$'s, *and* the newly introduced $\varepsilon_i$'s, (0.110) follows from the triangle inequality, and the last step follows since the $\bar{Z}_i$'s have the same distribution as the $Z_i$'s and the fact that the $Z_i$'s, $\bar{Z}_i$'s, and $\varepsilon_i$'s are all independent. Combining the above bound with (0.82) completes the proof of the theorem.

Example 0.7 above shows that deriving useful upper bounds on the Rademacher complexity can require considerable ingenuity, often relying on techniques tailored to the specific structure of the problem. Since the main goal of analyzing generalization error is to formalize the intuition that more data (i.e., larger $n$) leads to better generalization, we next examine how, under suitable conditions, the Rademacher complexity $\mathsf{R}_n(\mathscr{W})$ scales with $n$ and how it depends on the size of the model class $\mathscr{W}$.

### 0.4.4 Infinite model classes revisited

We revisit the idea of quantizing the model class $\mathscr{W}$ based on the empirical loss $L_n(w)$. This is made possible by the presence of the i.i.d. Rademacher random variables $\varepsilon_1, \varepsilon_2, \dots, \varepsilon_n$ in (0.101): if we condition on the dataset $Z_1, Z_2, \dots, Z_n$, we are still left with the randomness generated by the $\varepsilon_i$'s and can make use of metric entropy for an appropriately defined pseudometric that depends on $Z_1, Z_2, \dots, Z_n$. To that end, fix an outcome $z^n$ of the dataset and, for each $w, w' \in \mathscr{W}$, define the pseudometric

$$\rho_n(w, w') = \Big(\frac{1}{n}\sum_{i=1}^{n} |\ell(z_i, w) - \ell(z_i, w')|^2\Big)^{1/2}. \tag{0.112}$$

**Definition 0.13.** Given the pseudometric $\rho_n$, we define the *data-dependent* covering number and metric entropy as

$$N_n(\delta; \mathscr{W}, \rho_n) = \min\{N : \sup_{w\in\mathscr{W}} \min_{1\le j\le N} \rho_n(w, w_j) \le \delta \text{ for some } w_1, w_2, \dots, w_N \in \mathscr{W}\}, \tag{0.113}$$

$$H_n(\delta; \mathscr{W}, \rho_n) = \log N_n(\delta; \mathscr{W}, \rho_n). \tag{0.114}$$

While both $N_n$ and $H_n$ depend on $z^n$, in the following, we do not show this dependency explicitly to reduce notational clutter.

We now use the above definitions to obtain a useful upper bound on the empirical Rademacher complexity of learning problems. Fix an outcome of $\varepsilon_1, \varepsilon_2, \dots, \varepsilon_n$. Then, for any two $w, w' \in \mathscr{W}$, we have

$$\Big|\frac{1}{n}\sum_{i=1}^{n}\varepsilon_i \ell(z_i, w)\Big| - \Big|\frac{1}{n}\sum_{i=1}^{n}\varepsilon_i \ell(z_i, w')\Big| \le \Big|\frac{1}{n}\sum_{i=1}^{n}\varepsilon_i (\ell(z_i, w) - \ell(z_i, w'))\Big| \tag{0.115}$$

$$\le \frac{1}{n}\sum_{i=1}^{n}|\ell(z_i, w) - \ell(z_i, w')| \tag{0.116}$$

$$\le \rho_n(w, w'), \tag{0.117}$$

where (0.115) follows from the triangle inequality, (0.116) follows from Jensen's inequality, and the final step follows from the definition of $\rho_n$ and Jensen's inequality.

For any minimal $\delta$-net $w_1, w_2, \dots, w_N$ of $\mathscr{W}$ with $N = N_n(\delta; \mathscr{W}, \rho_n)$, we have

$$\begin{aligned}\frac{1}{n}\sup_{w\in\mathscr{W}}\Big|\sum_{i=1}^{n}\varepsilon_i \ell(z_i, w)\Big| &\le \sup_{w\in\mathscr{W}}\min_{1\le j\le N}\rho_n(w, w_j) + \frac{1}{n}\max_{1\le j\le N}\Big|\sum_{i=1}^{n}\varepsilon_i \ell(z_i, w_j)\Big| \\ &\le \delta + \frac{1}{n}\max_{1\le j\le N}\Big|\sum_{i=1}^{n}\varepsilon_i \ell(z_i, w_j)\Big|.\end{aligned}$$

Taking expectations only with respect to $\varepsilon^n$ gives

$$\bar{\mathsf{R}}_n(\mathscr{W}, z^n) \le \delta + \frac{1}{n}\,\mathsf{E}_{\varepsilon^n}\Big(\max_{1\le j\le N}\Big|\sum_{i=1}^n \varepsilon_i \ell(z_i, w_j)\Big|\Big).$$

Note that the second term on the right-hand side involves the expectation of the maximum of $N = N_n(\delta; \mathscr{W}, \rho_n)$ random variables. Since the $\varepsilon_i$'s are i.i.d. 1-subgaussian random variables, it follows that each random variable in the maximum is $\sigma_j^2$-subgaussian, with $\sigma_j^2 = \sum_{i=1}^n \ell^2(z_i, w_j)$. Hence we have established the following.

**Lemma 0.10.** Consider a model class $\mathscr{W}$ and a fixed dataset outcome $z^n$. Let $\rho_n$ be the pseudometric on $\mathscr{W}$ as defined in (0.112), and let $(w_1, \dots, w_N)$ be a $\delta$-net of $\mathscr{W}$ with $N = N_n(\delta; \mathscr{W}, \rho_n)$. Then,

$$\bar{\mathsf{R}}_n(\mathscr{W}, z^n) \le \delta + \frac{3}{n}\max_{1\le j\le N}\sigma_j\sqrt{H_n(\delta; \mathscr{W}, \rho_n)}. \tag{0.118}$$

To proceed towards a bound on $\mathsf{E}(\Delta_n(\mathscr{W}))$, we need additional assumptions on $\ell$ and $\mathscr{W}$. For example, we may assume the following:

**Assumption 1.** The loss function $\ell$ takes values in a finite interval $[0, b]$.

**Assumption 2.** For any $\delta > 0$,

$$H(\delta; \mathscr{W}) = \sup_{n\ge 1}\sup_{z^n} H_n(\delta; \mathscr{W}, \rho_n) < \infty. \tag{0.119}$$

The first of these is self-explanatory. The second, though, requires some motivation. Essentially, it states that: (i) for each sample size $n$ and each outcome $z^n$ of the dataset, we can approximate the infinite class $\mathscr{W}$ by a *data-dependent* finite set $\hat{\mathscr{W}}_n$ in the sense that the empirical loss of any $w \in \mathscr{W}$ on $z^n$ can be approximated by the empirical loss of some $\hat{w} \in \hat{\mathscr{W}}_n$; and (ii) the size of $\hat{\mathscr{W}}_n$ can be bounded uniformly in $n$ and in the outcome $z^n$ of the dataset. While these appear to be very strong requirements, in the following section we present an important learning problem where it holds.

With these two assumptions in place, we see that the inequality

$$\bar{\mathsf{R}}_n(\mathscr{W}, z^n) \le \inf_{\delta>0}\Big(\delta + 3b\sqrt{\frac{H(\delta; \mathscr{W})}{n}}\Big) \tag{0.120}$$

holds for *every* outcome $z^n$ of the dataset. We can then use the law of iterated expectation and the fact that $Z_1, Z_2, \dots, Z_n$ are independent of $\varepsilon_1, \varepsilon_2, \dots, \varepsilon_n$ to obtain

$$\mathsf{R}_n(\mathscr{W}) \le \inf_{\delta>0}\Big(\delta + 3b\sqrt{\frac{H(\delta; \mathscr{W})}{n}}\Big). \tag{0.121}$$

Finally, recalling Theorem 0.3, we arrive at the following bound.

**Theorem 0.4.** Let $(\mathscr{W}, \rho)$ be a metric space, and let $\delta > 0$. Under Assumptions (1) and (2) above, we have

$$\mathsf{E}(\Delta_n(\mathscr{W})) \leq \inf_{\delta>0} \left( 2\delta + 6b\sqrt{\frac{H(\delta;\mathscr{W})}{n}} \right). \tag{0.122}$$

Combining this with (0.65), we obtain the following bound on the expected excess risk of ERM.

**Corollary 0.1.** Consider the ERM algorithm, and let $\delta > 0$. Then,

$$\mathsf{E}\left(L(w_{\mathrm{ERM}}) - \inf_{w\in\mathscr{W}} L(w)\right) \leq \inf_{\delta>0} \left( 2\delta + 6b\sqrt{\frac{H(\delta;\mathscr{W})}{n}} \right). \tag{0.123}$$

The significance of the bounds in (0.122) and (0.123) is that, under suitable assumptions on $\mathscr{W}$ and $\ell$, we can upper-bound the excess risk of ERM in a *distribution-free* manner, i.e., independent of the distribution of $Z_i$.

**Remark.** In some cases, such as the binary classification example discussed in the next section, it is possible to upper-bound $H(\delta;\mathscr{W})$ defined in (0.119) by

$$H(\delta;\mathscr{W}) \leq C(\mathscr{W}) \log\frac{1}{\delta}, \quad \delta > 0 \tag{0.124}$$

for some constant $C(\mathscr{W})$. By analogy with the bounds on the metric entropy of the unit ball in $\mathbb{R}^d$, we may think of $C(\mathscr{W})$ as a measure of the "effective dimension" of $\mathscr{W}$ in the learning problem with loss function $\ell$. We can then take $\delta = 1/n$ in (0.122) and (0.123) to obtain bounds of the form

$$\mathsf{E}\left(L(w_{\mathrm{ERM}}) - \inf_{w\in\mathscr{W}} L(w)\right) \leq \mathrm{const} \cdot \sqrt{\frac{C(\mathscr{W})\log n}{n}}. \tag{0.125}$$

A more sophisticated analysis using a technique called *chaining* can be used to remove the $\log n$ factor in the above inequality, as suggested in (0.69). There is an extensive literature devoted to these issues; see, for example, Chapter 13 of (Boucheron, Lugosi, and Massart 2013).

### 0.4.5 Binary Classification and VC Dimension

A simple but important case for which the assumptions in Section 0.4.4 hold is binary classification. Here the data space is $\mathscr{Z} = \mathscr{X} \times \{0, 1\}$ and the model class consists of a family of classifiers $\phi(x, w)$ indexed by $w \in \mathscr{W}$. For example, when $\mathscr{X} = \mathbb{R}^K$, the classifiers may be linear or use neural networks. As before, we assume the $0-1$ loss function:

$$\ell(z, w) = \ell((x, y), w) = \mathbb{1}_{\{y\neq\phi(x,w)\}}.$$

We are given a dataset $(X_1, Y_1), \ldots, (X_n, Y_n)$ with the $X_i$s taking values in the feature set $\mathscr{X}$ and the $Y_i$s taking values in $\{0, 1\}$ and we wish to find a classifier that minimizes the empirical risk. In the case of linear classifiers, the ERM algorithm amounts to finding a hyperplane $\mathscr{A}_{\mathbf{w}} = \{\mathbf{x} \in \mathbb{R}^K : \mathbf{w}^T\mathbf{x} = 0\}$ that best *separates* the dataset in the sense that the number of misclassified pairs $(X_i, Y_i)$ with $\phi(X_i, \mathbf{w}) \neq Y_i$ is minimized (Figure 0.2).

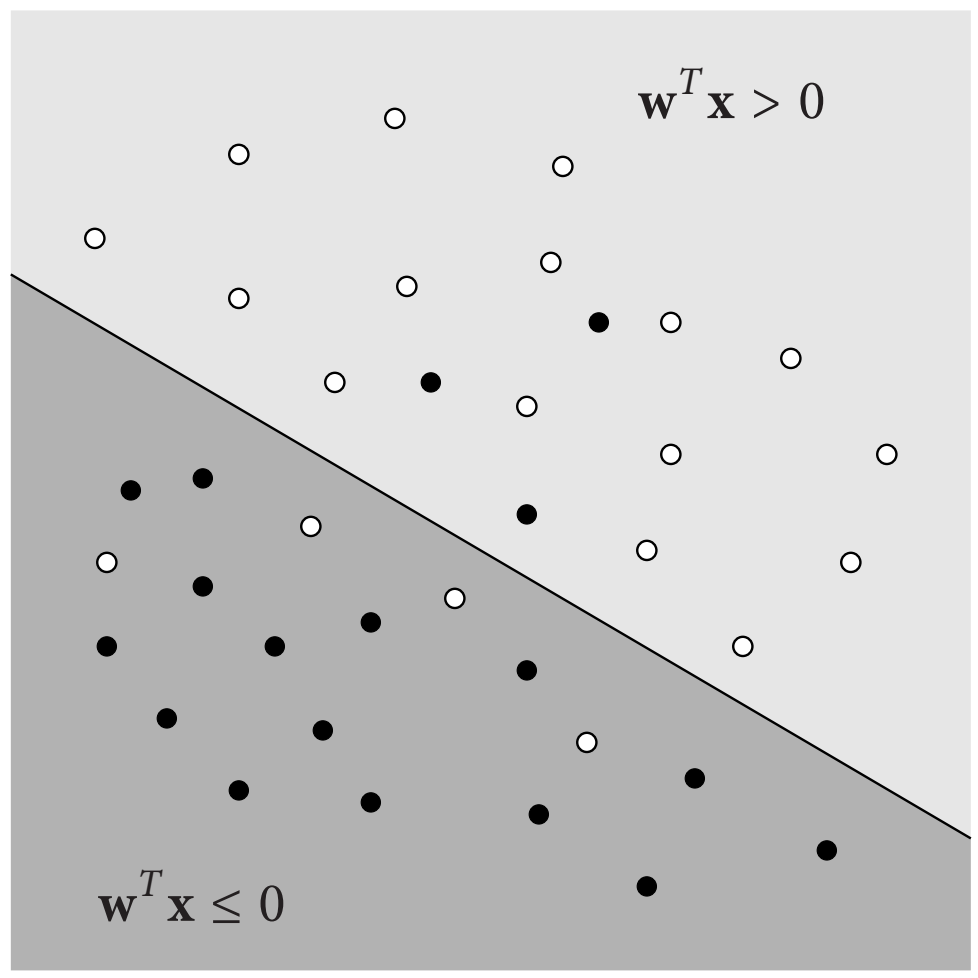

**Figure 0.2.** Linear classifier.

In this setup, Assumption 1 holds with $b = 1$. To show that Assumption 2 holds, we first introduce the following classical concepts from learning theory.

**Definition 0.14.** Let $\mathscr{C}$ be a class of measurable subsets of $\mathscr{X}$. For each $n = 1, 2, \ldots$, the $n$th *shatter coefficient* of $\mathscr{C}$ is defined as

$$S_n(\mathscr{C}) = \max_{x_1, x_2, \ldots, x_n \in \mathscr{X}} \left|\{\left(\mathbb{1}_{\{x_1 \in \mathscr{A}\}}, \ldots, \mathbb{1}_{\{x_n \in \mathscr{A}\}}\right) : \mathscr{A} \in \mathscr{C}\}\right|. \tag{0.126}$$

It is clear that $S_n(\mathscr{C}) \le 2^n$.

**Definition 0.15.** The *Vapnik–Chervonenkis (VC) dimension* of $\mathscr{C}$ is the largest integer $n \ge 1$ such that $S_n(\mathscr{C}) = 2^n$ (if $S_n(\mathscr{C}) = 2^n$ for all $n$, then we say that $\mathscr{C}$ has infinite VC dimension). If $\mathscr{C}$ has finite VC dimension, then we say that it is a *VC class* or that it has the VC property.

**Examples of VC classes.**

1. The class of all half-spaces of $\mathbb{R}^d$, i.e., sets of the form $\mathscr{C}_{\mathbf{w}} = \{\mathbf{x} \in \mathbb{R}^d : \mathbf{w}^T\mathbf{x} \ge 0\}$ with $\mathbf{w} \in \mathbb{R}^d$. This class has VC dimension $d$;

2. Consider the class of sets of the form

$$\mathscr{C}_{\mathbf{w},h} = \{x \in \mathscr{X} : \mathbf{w}^T\phi(x) + h(x) \ge 0\}, \tag{0.127}$$

where parameter vector $\mathbf{w}$ is a vector in $\mathbb{R}^m$, $\phi(x) = \left[\phi_1(x)\ \phi_2(\mathbf{x})\ \cdots\ \phi_m(x)\right]^T$ is a vector-valued function from an arbitrary set $\mathscr{X}$ into $\mathbb{R}^m$, whose coordinate functions $\phi_1, \phi_2, \ldots, \phi_m : \mathcal{X} \to \mathbb{R}$ are linearly independent, and $h : \mathcal{X} \to \mathbb{R}$ is an arbitrary function.

This class generalizes linear classifiers based on half-spaces: the mapping $\phi(x)$ may be interpreted as an $m$-dimensional vector of nonlinear *features* of $\mathcal{X}$. It can be shown that this class has VC dimension $m$ (Dudley 1978).

A key result in the theory of VC classes is the following estimate on the shatter coefficients, derived independently (and in different contexts) by Vapnik and Chervonenkis (1971), Sauer (1972), and Shelah (1972).

**Lemma 0.11.** Let $\mathcal{C}$ be a VC class with VC dimension $D$. Then its shatter coefficients for $n \geq D$ grow polynomially in $n$ as

$$S_n(\mathcal{C}) \leq \left(\frac{ne}{D}\right)^D. \tag{0.128}$$

We now return to the binary classification problem. Each classifier $\phi(x, w)$ induces a subset of $\mathcal{X}$, namely $\mathcal{C}_w = \{x \in \mathcal{X} : \phi(x, w) = 1\}$. We show that Assumption 2 from the previous section holds whenever these sets form a VC class.

**Lemma 0.12.** Suppose the family of binary classifiers $\{\phi(x, w) : w \in \mathcal{W}\}$ is such that the associated class of sets $\mathcal{C}_{\mathcal{W}} = \{\mathcal{C}_w : w \in \mathcal{W}\}$ is a Vapnik–Chervonenkis class with VC dimension $D$. Then, for every $\delta > 0$,

$$H(\delta; \mathcal{W}) = \sup_{n \geq 1} \sup_{z^n} H_n(\delta; \mathcal{W}, \rho_n) \leq c \cdot D \log \frac{1}{\delta}, \tag{0.129}$$

where $\rho_n$ is as defined in (0.112) and $c > 0$ is a universal constant.

**Proof.** Fix a dataset outcome $(x_1, y_1), \ldots, (x_n, y_n)$. Then, for any $w, w' \in \mathcal{W}$, we have

$$\rho_n(w, w') = \left(\frac{1}{n}\sum_{i=1}^{n} \left|\mathbb{1}_{\{y_i \neq \phi(x_i, w)\}} - \mathbb{1}_{\{y_i \neq \phi(x_i, w')\}}\right|^2\right)^{1/2}. \tag{0.130}$$

Since we are considering the Hamming metric, for each $i$,

$$\left|\mathbb{1}_{\{y_i \neq \phi(x_i, w)\}} - \mathbb{1}_{\{y_i \neq \phi(x_i, w')\}}\right|^2 = \mathbb{1}_{\{\phi(x_i, w) \neq \phi(x_i, w')\}}. \tag{0.131}$$

Each classifier $\phi(\cdot, w)$ defines a subset of $\mathcal{X}$, $\mathcal{C}_w = \{x \in \mathcal{X} : \phi(x, w) = 1\}$, and we can verify that, for any $x$, $\mathbb{1}_{\{\phi(x,w) \neq \phi(x,w')\}} = 1$ if and only if $x$ belongs to the symmetric difference of $\mathcal{C}_w$ and $\mathcal{C}_{w'}$,

$$\mathcal{C}_w \triangle \mathcal{C}_{w'} = (\mathcal{C}_w \cup \mathcal{C}_{w'}) \setminus (\mathcal{C}_w \cap \mathcal{C}_{w'}).$$

Hence, we have

$$\rho_n(w, w') = \left(\frac{1}{n}\sum_{i=1}^{n} \mathbb{1}_{\{x_i \in \mathcal{C}_w \triangle \mathcal{C}_{w'}\}}\right)^{1/2}, \tag{0.132}$$

which depends on $w$ and $w'$ through the sets $\mathcal{C}_w$ and $\mathcal{C}_{w'}$ and involves only $x^n$, but not $y^n$. Let $P_n$ be the empirical probability measure on $\mathscr{X}$ induced by $x^n$ that assigns mass $1/n$ to each $x_i$ and zero to all other $x \in \mathscr{X}$. Then $\rho_n(w, w')$ can alternatively be expressed as

$$\rho_n(w, w') = \sqrt{P_n(\mathscr{C}_w \triangle \mathscr{C}_{w'})},$$

i.e., as the square root of the empirical probability of $\mathscr{C}_w \triangle \mathscr{C}_{w'}$ under $P_n$.

It is not difficult to see that any probability measure $Q$ on $\mathscr{X}$ induces a pseudometric on the collection of all measurable subsets of $\mathscr{X}$, defined as $\rho_Q(\mathscr{A}, \mathscr{B}) = \sqrt{Q(\mathscr{A} \triangle \mathscr{B})}$. Hence, the relationship between $\rho_n$ and $\rho_{P_n}$ can be expressed as

$$\rho_n(w, w') = \rho_{P_n}(\mathscr{C}_w, \mathscr{C}_{w'}). \tag{0.133}$$

We can define its covering number $N(\delta; \mathscr{C}_{\mathscr{W}}, \rho_Q)$ and metric entropy $H(\delta; \mathscr{C}_{\mathscr{W}}, \rho_Q)$ with respect to $\rho_Q$. Using Lemma 0.11 together with a random coding argument, it can be shown that, for any probability measure Q and any $0 < \delta < 1$,

$$H(\delta; \mathscr{C}_{\mathscr{W}}, \rho_Q) \le c \cdot D \log \frac{1}{\delta}, \tag{0.134}$$

where $c > 0$ is a universal constant that is independent of both $Q$ and $\mathscr{C}_{\mathscr{W}}$. The detailed steps are carried out in Problem 0.15.

By the assumption that the class of sets $\mathscr{C}_{\mathscr{W}} = \{\mathscr{C}_w : w \in \mathscr{W}\}$ for the family of classifiers is a VC class with VC dimension $D$. From (0.133) and (0.134), it follows that for $0 < \delta \le 1$,

$$H(\delta; \mathscr{W}) \le \sup_Q H(\delta; \mathscr{C}_{\mathscr{W}}, \rho_Q) \le c \cdot D \log \frac{1}{\delta}.$$

This completes the proof of the lemma.

Combining Lemma 0.12 with the bound in Corollary 0.1, we obtain the following bound on the worst-case generalization error of ERM for binary classification.

**Theorem 0.5.** Let $\mathscr{W}$ be a family of binary classifiers, such that the induced class of sets $\mathscr{C}_{\mathscr{W}}$ is a VC class with VC dimension $D$. Then the excess risk of the empirical risk minimizer satisfies

$$\mathsf{E}\left(L(w_{\text{ERM}}) - \inf_{w \in \mathscr{W}} L(w)\right) \le c \cdot \sqrt{\frac{D}{n}}. \tag{0.135}$$

**Remark.** We can give a clear operational interpretation of the VC dimension of $\mathscr{C}_{\mathscr{W}}$. For each $n$, the shatter coefficient $S_n(\mathscr{C}_{\mathscr{W}})$ is the maximum number of distinct binary labelings that can be realized on any $n$ points $x_1, x_2, \ldots, x_n \in \mathscr{X}$ by classifiers of the form $\phi(x, w)$. When the sample size $n$ is sufficiently large—specifically, when $n \ge D$, where $D$ is the VC dimension of $\mathscr{C}_{\mathscr{W}}$—the family $\{\phi(x, w)\}$ can generate at most $O(n^D)$ distinct labelings on any $n$ points of $\mathscr{X}$.

This polynomial growth constraint limits the capacity of the classifier class and prevents overfitting when the training sample is large enough. In particular, it rules out the possibility that the class is so rich that, for every $n$, there exists a classifier that perfectly fits any assignment of labels to the $n$ training samples.

## 0.5 BOUNDS ON EXPECTED GENERALIZATION ERROR

While the bounds discussed in the previous section are motivated by the ERM algorithm, they can be overly conservative for other learning methods. To provide guarantees on generalization beyond ERM, we derive bounds on the expected generalization error $\mathsf{E}\left(L(W) - L_n(W)\right)$, where the expectation is taken over both the training data $Z^n$ and the algorithm's output $W$. These bounds are obtained from a single bound expressed in terms of the mutual information between $Z^n$ and $W$.

**Notation.** For convenience in this and the following sections, we use the natural logarithm in defining relative entropy $D$ and mutual information $I$, rather than the base 2 logarithm adopted in most of the text.

### 0.5.1 Bounds using mutual information

The empirical risk $L_n(W)$ is a random variable that depends on the training dataset $Z^n$. To make this dependence explicit, we define the function

$$\mathcal{L}(W, Z^n) = \frac{1}{n}\sum_{i=1}^{n} \ell(Z_i, W). \tag{0.136}$$

Recalling our discussion of symmetrization in Section 0.4.3, let $\bar{Z}^n$ be an i.i.d. sequence of auxiliary random variables independent of $Z^n$, and hence independent of $W$. We can then express the expected generalization error in terms of $\mathcal{L}$ as

$$\mathsf{E}\left(L(W) - L_n(W)\right) = \mathsf{E}\left(\mathcal{L}(W, \bar{Z}^n) - \mathcal{L}(W, Z^n)\right). \tag{0.137}$$

The right-hand side involves three random objects: $W$, $Z^n$, and $\bar{Z}^n$. The sequences $Z^n$ and $\bar{Z}^n$ are i.i.d. and take values in $\mathscr{Z}^n$; the conditional distribution of $W$ given $Z^n$ is induced by the learning algorithm; and $\bar{Z}^n$ is independent of $(W, Z^n)$. To show that the expected generalization error is small, it is natural to focus on regimes in which $W$ does not depend too strongly on the training data $Z^n$. This dependence can be quantified using the mutual information between $W$ and $Z^n$.

To make this intuition precise, consider a pair of random variables $(U, V) \sim p_{U,V}(u, v)$, with $(u, v) \in \mathscr{U} \times \mathscr{V}$. Let $\bar{U}$ and $\bar{V}$ be independent copies of $U$ and $V$, with joint distribution $p_{\bar{U},\bar{V}}(u, v) = p_U(u)p_V(v)$. For an arbitrary function $\phi : \mathscr{U} \times \mathscr{V} \to \mathbb{R}$, we are interested in controlling the absolute difference between $\mathsf{E}(\phi(U, V))$ and $\mathsf{E}(\phi(\bar{U}, \bar{V}))$ in terms of the mutual information.

**Lemma 0.13.** Suppose that, for every $u \in \mathscr{U}$, the random variable $\phi(u, V)$ is $\sigma^2$-subgaussian under $p_V(v)$. Then

$$\left|\mathsf{E}(\phi(U, V)) - \mathsf{E}(\phi(\bar{U}, \bar{V}))\right| \le \sqrt{2\sigma^2 I(U; V)}. \tag{0.138}$$

**Proof.** Recall the variational representation of the relative entropy: for any two probability distributions $p, q$ on a common set $\mathcal{X}$,

$$D(p\|q) = \sup_{\psi}\Big( \mathsf{E}_p(\psi(X)) - \ln \mathsf{E}_q\left(e^{\psi(X)}\right)\Big), \tag{0.139}$$

where the supremum is over all functions $\psi : \mathcal{X} \to \mathbb{R}$, such that $\mathsf{E}_q\left(e^{\psi(X)}\right) < \infty$. For any $u \in \mathcal{U}$ and $\lambda \in \mathbb{R}$, applying (0.139) to $p = p_{V|U}(v|u)$ and $q = p_V(v)$ on $\mathcal{X} = \mathcal{V}$ with $\psi(v) = \lambda\phi(u, v)$, we obtain

$$\begin{aligned} D(p(v|u)\|p(v)) &\geq \mathsf{E}(\lambda\phi(u, V)|U = u) - \ln \mathsf{E}\left(\mathrm{e}^{\lambda\phi(u,V)}\right) \\ &\geq \lambda\big( \mathsf{E}(\phi(u, V)|U = u) - \mathsf{E}(\phi(u, V))\big) - \frac{\lambda^2\sigma^2}{2}, \end{aligned} \tag{0.140}$$

where the second step follows from the subgaussian assumption on $\phi(u, V)$,

$$\ln \mathsf{E}\left(\mathrm{e}^{\lambda(\phi(u,V)-\mathsf{E}(\phi(u,V)))}\right) \leq \frac{\lambda^2\sigma^2}{2} \quad \text{for every } \lambda \in \mathbb{R}. \tag{0.141}$$

Optimizing the right-hand side of (0.142) over $\lambda$ and rearranging, we obtain

$$\big|\mathsf{E}(\phi(u, V)|U = u) - \mathsf{E}(\phi(u, V))\big| \leq \sqrt{2\sigma^2 D(p(v|u)\|p(v))}. \tag{0.142}$$

Then,

$$\begin{aligned} \big|\mathsf{E}(\phi(U, V)) - \mathsf{E}(\phi(\bar{U}, \bar{V}))\big| &= \big|\mathsf{E}_U\big( \mathsf{E}(\phi(U, V) - \phi(U, \bar{V})|U)\big)\big| && (0.143) \\ &\leq \mathsf{E}_U\Big(\big|\mathsf{E}\left(\phi(U, V)|U\right) - \mathsf{E}\left(\phi(U, \bar{V})|U\right)\big|\Big) && (0.144) \\ &\leq \mathsf{E}_U\Big(\sqrt{2\sigma^2 D(p(v|u)\|p(v))}\Big) && (0.145) \\ &\leq \sqrt{2\sigma^2 D(p(u, v)\|p(u)p(v))}, && (0.146) \end{aligned}$$

where (0.144) follows from Jensen's inequality, (0.145) follows from (0.142), and the last step follows again by Jensen's inequality. To complete the proof, note that $I(U; V) = D(p(u, v)\|p(u)p(v))$.

We can now obtain our first information-theoretic generalization bound.

**Theorem 0.6.** Suppose that $\ell(Z, w)$ is $\sigma^2$-subgaussian for every $w \in \mathcal{W}$. Then

$$\big|\mathsf{E}\left(L(W) - L_n(W)\right)\big| \leq \sqrt{\frac{2\sigma^2}{n} I(W; Z^n)}. \tag{0.147}$$

**Proof.** Since the $Z_i$'s are independent, the subgaussian assumption on the loss implies that, for every $w \in \mathcal{W}$, the empirical risk $\mathcal{L}(w, Z^n)$ is $(\sigma^2/n)$-subgaussian. Applying Lemma 0.13 with $U = W$, $V = Z^n$, and $\phi(u, v) = \mathcal{L}(w, z^n)$ then yields the inequality (0.147).

**Corollary 0.2.** If $\mathscr{W}$ is finite, then for any learning algorithm,

$$\left|\mathsf{E}\left(L(W) - L_n(W)\right)\right| \leq \sqrt{\frac{2\sigma^2}{n} \ln |\mathscr{W}|}. \tag{0.148}$$

**Proof.** The result follows directly from Theorem 0.6. Indeed, when $\mathscr{W}$ is finite,

$$I(W; Z^n) = H(W) - H(W|Z^n) \leq H(W) \leq \ln |\mathscr{W}|. \tag{0.149}$$

Substituting this bound into (0.147) yields the desired inequality.

We can still obtain nontrivial bounds for infinite $\mathscr{W}$ provided the learning algorithm is relatively insensitive to changes in any single sample $Z_i$.

**Theorem 0.7.** Suppose that there exists $\epsilon > 0$ such that, for any pair of datasets $z^n$ and $\bar{z}^n$ differing in at most one coordinate $z_i$,

$$D(p(w|z^n) \| p(w|\bar{z}^n)) \leq \epsilon. \tag{0.150}$$

Then $I(W; Z^n) \leq n\epsilon$. Consequently,

$$\left|\mathsf{E}(L(W) - L_n(W))\right| \leq \sqrt{2\sigma^2 \epsilon}. \tag{0.151}$$

**Proof.** From the convexity of relative entropy, we have

$$D(p(w|z^n) \| p(w|z^{i-1}, z_{i+1}^n)) \leq \mathsf{E}_{\bar{Z}_i} \left( D(p(w|z^n) \| p(w|z^{i-1}, \bar{Z}_i, z_{i+1}^n)) \right), \tag{0.152}$$

where $\bar{Z}_i$ has the same distribution as $Z_i$. Combining this with (0.150), we can upper bound the conditional mutual information as

$$I(W; Z_i | Z^{i-1}, Z_{i+1}^n) = \mathsf{E}_{Z^n} \left( D(p(w|Z^n) \| p(w|Z^{i-1}, Z_{i+1}^n)) \right) \leq \epsilon. \tag{0.153}$$

Moreover,

$$\begin{aligned} I(W; Z^n) &= \sum_{i=1}^n I(W; Z_i | Z^{i-1}) && (0.154) \\ &= \sum_{i=1}^n I(W, Z^{i-1}; Z_i) && (0.155) \\ &\leq \sum_{i=1}^n I(W, Z^{i-1}, Z_{i+1}^n; Z_i) && (0.156) \\ &= \sum_{i=1}^n I(W; Z_i | Z^{i-1}, Z_{i+1}^n), && (0.157) \end{aligned}$$

where (0.155) and (0.157) follow since the $Z_i$'s are independent. (The quantity in (0.157) is referred to as the *erasure information* between $W$ and $Z^n$ (Verdú and Weissman 2008).) Combining (0.157) and (0.153), it follows that $I(W; Z^n) \leq n\epsilon$. The rest of the proof follows from Theorem 0.6.

### 0.5.2 The Gibbs algorithm

Except in special cases—such as when $\mathscr{W}$ is finite—the information-theoretic generalization bound of Theorem 0.6 is not helpful for deterministic learning algorithms, since the mutual information $I(W; Z^n)$ is generally infinite. In this section, we study a stochastic relaxation of ERM known as the *Gibbs algorithm*. Rather than outputting a single deterministic model, the Gibbs algorithm produces a random variable $W$ that, with high probability, attains small empirical loss $L_n(W)$.

We assume that $\mathscr{W} \subset \mathbb{R}^d$ for some $d$, that is, the model class admits a finite-dimensional parametrization by a $d$-dimensional vector $\mathbf{w} \in \mathbb{R}^d$, as in linear classifiers and neural nets. Fix a data-independent pdf $\pi(w)$ supported on $\mathscr{W}$, which, in this section and in the next, is *not* necessarily the marginal pdf of $W$ under the joint pdf $f(w, z^n) = f(z^n)f(w|z^n)$. Also fix an "inverse temperature parameter" $\beta > 0$. The Gibbs algorithm is then defined by the conditional pdf

$$f_\beta(w|z^n) = \frac{\exp\big(-\beta\mathcal{L}(w, z^n)\big)\pi(w)}{\int_{\mathscr{W}} \exp\big(-\beta\mathcal{L}(w', z^n)\big)\pi(w')\,dw'}. \tag{0.158}$$

Under mild regularity assumptions on the loss $\ell$ and the pdf $\pi$, the ERM solution is recovered in the limit as $\beta \to \infty$; see Problem 0.17.

**Theorem 0.8.** Suppose the loss function $\ell$ takes values in $[0, 1]$. Then the expected generalization error of the Gibbs algorithm defined in (0.158) satisfies

$$\big|\mathsf{E}\big(L(W) - L_n(W)\big)\big| \le \frac{\beta}{2n}. \tag{0.159}$$

**Proof.** Consider two datasets $z^n$ and $\bar{z}^n$ that differ in exactly one coordinate and define

$$\Delta_{\mathcal{L}}(W) = \beta\big(\mathcal{L}(W, z^n) - \mathcal{L}(W, \bar{z}^n)\big). \tag{0.160}$$

Since the loss function $\ell \in [0, 1]$, $\Delta_{\mathcal{L}}(W) \in [-\beta/n, \beta/n]$. Hence, by Hoeffding's Lemma 0.1, $\Delta_{\mathcal{L}}(W)$ is $\beta^2/n^2$-subgaussian. Also define

$$A(z^n) = \ln\Big(\int_{\mathscr{W}} \exp\big(-\beta\mathcal{L}(w', z^n)\big)\pi(w')\,dw'\Big). \tag{0.161}$$

Now, consider

$$\begin{aligned}
D\big(f_\beta(w|z^n)\|f_\beta(w|\bar{z}^n)\big) &= -\mathsf{E}_{f_\beta(w|z^n)}\big(\Delta_{\mathcal{L}}(W)\big) + A(\bar{z}^n) - A(z^n) && (0.162)\\
&= -\mathsf{E}_{f_\beta(w|z^n)}\big(\Delta_{\mathcal{L}}(W)\big) + \ln \mathsf{E}_{f_\beta(w|z^n)}\Big(e^{\Delta_{\mathcal{L}}(W)}\Big) && (0.163)\\
&\le -\mathsf{E}_{f_\beta(w|z^n)}\big(\Delta_{\mathcal{L}}(W)\big) + \mathsf{E}_{f_\beta(w|z^n)}\big(\Delta_{\mathcal{L}}(W)\big) + \frac{\beta^2}{2n^2} && (0.164)\\
&= \frac{\beta^2}{2n^2}, && (0.165)
\end{aligned}$$

where (0.162) follows from the expression for the Gibbs algorithm (0.158), and (0.164) follows since $\Delta_{\mathcal{L}}(W)$ is $\beta^2/n^2$-subgaussian (definition (0.12) with $\lambda = 1$). To justify (0.163), we note that

$$
\begin{aligned}
A(\bar{z}^n) - A(z^n) &= \ln e^{A(\bar{z}^n) - A(z^n)} && (0.166)\\
&= \ln \frac{f_\beta(w|z^n) \exp\big(-\beta\mathcal{L}(w, \bar{z}^n)\big)\pi(w)}{f_\beta(w|\bar{z}^n) \exp\big(-\beta\mathcal{L}(w, z^n)\big)\pi(w)} \quad \text{for every } \ w \in \mathcal{W} && (0.167)\\
&= \ln \mathsf{E}_{f_\beta(w|\bar{z}^n)}\left(\frac{f_\beta(W|z^n) \exp\big(-\beta\mathcal{L}(W, \bar{z}^n)\big)\pi(W)}{f_\beta(W|\bar{z}^n) \exp\big(-\beta\mathcal{L}(W, z^n)\big)\pi(W)}\right) && (0.168)\\
&= \ln \mathsf{E}_{f_\beta(w|\bar{z}^n)}\left(\frac{f_\beta(W|z^n)}{f_\beta(W|\bar{z}^n)} e^{\Delta_{\mathcal{L}}(W)}\right) && (0.169)\\
&= \ln \mathsf{E}_{f_\beta(w|z^n)}\left(e^{\Delta_{\mathcal{L}}(W)}\right), && (0.170)
\end{aligned}
$$

where (0.167) follows from the definition of $A(\mathbf{z})$ and the Gibbs algorithm (0.158). Therefore, condition (0.150) in Theorem 0.7 holds with $\epsilon = \beta^2/2n^2$, and the mutual information between $Z^n$ and $W$ under the Gibbs algorithm satisfies $I(W; Z^n) \leq \beta^2/2n$. Moreover, since the loss $\ell$ takes values in $[0, 1]$, it is $1/2$-subgaussian, hence the subgaussian assumption on the loss holds with $\sigma = 1/2$. Combining the results completes the proof.

For a general subgaussian loss, we can establish the following bound.

**Theorem 0.9.** Suppose that $\ell(w, Z)$ is $\sigma^2$-subgaussian for every $w \in \mathcal{W}$. Then the expected risk of the Gibbs algorithm (0.158) satisfies

$$
\mathsf{E}(L(W)) \leq -\frac{1}{\beta} \mathsf{E}_{Z^n}\left(\ln \mathsf{E}_{\tilde{W}\sim\pi}\left(e^{-\beta\mathcal{L}(\tilde{W}, Z^n)}\right)\right) + \frac{\beta\sigma^2}{2n}, \tag{0.171}
$$

where the expectation on the left-hand side is with respect to $f_{Z^n}(z^n) f_{W|Z^n}(w|z^n)$ and the inner expectation on the RHS is over $\tilde{W} \sim \pi$ independent of $Z^n$.

**Proof.** Consider any learning algorithm for which $I(W; Z^n) < \infty$. We begin with the bound from Theorem 0.6:

$$
\left|\mathsf{E}\left(L(W) - \mathcal{L}(W, Z^n)\right)\right| \leq \sqrt{\frac{2\sigma^2}{n} I(W; Z^n)}\,.
$$

Removing the absolute value and rearranging yields the bound on the expected risk:

$$
\mathsf{E}[L(W)] \leq \mathsf{E}[\mathcal{L}(W, Z^n)] + \sqrt{\frac{2\sigma^2}{n} I(W; Z^n)}\,. \tag{0.172}
$$

We then linearize the square root using the arithmetic–geometric mean inequality $2ab \leq a^2 + b^2$, valid for any $a, b \in \mathbb{R}$. Set

$$
a^2 = \frac{I(W; Z^n)}{\beta} \quad \text{and} \quad b^2 = \frac{\beta\sigma^2}{2n}.
$$

Then,

$$2ab = 2\sqrt{\frac{I(W;Z^n)}{\beta}\cdot\frac{\beta\sigma^2}{2n}} = \sqrt{\frac{2\sigma^2 I(W;Z^n)}{n}},$$

and

$$\sqrt{\frac{2\sigma^2 I(W;Z^n)}{n}} \le \frac{I(W;Z^n)}{\beta} + \frac{\beta\sigma^2}{2n}.$$

Substituting into (0.172) gives a bound that is linear in the mutual information term, valid for any learning algorithm and any $\beta > 0$:

$$\mathsf{E}(L(W)) \le \mathsf{E}(\mathcal{L}(W, Z^n)) + \frac{I(W;Z^n)}{\beta} + \frac{\beta\sigma^2}{2n}. \tag{0.173}$$

The rest of the proof identifies the algorithm that minimizes a bound on the first two terms. Using the mutual information identity:

$$I(W;Z^n) = \mathsf{E}\left(D(f(w|Z^n)\|\pi(w))\right) - D(f_W(w)\|\pi(w)). \tag{0.174}$$

Then, from the nonnegativity of relative entropy, we obtain the bound:

$$I(W;Z^n) \le \mathsf{E}\left(D(f(w|Z^n)\|\pi(w))\right).$$

Hence,

$$\begin{aligned}
&\mathsf{E}\left(\mathcal{L}(W, Z^n)\right) + \frac{1}{\beta} I(W;Z^n) \\
&\qquad \le \mathsf{E}\left(\mathcal{L}(W, Z^n)\right) + \frac{1}{\beta}\,\mathsf{E}\left(D(f(w|Z^n)\|\pi(w))\right) && (0.175)\\
&\qquad = \mathsf{E}_{Z^n}\Big(\,\mathsf{E}_{W\sim f(w|Z^n)}\left(\mathcal{L}(W, Z^n)\right) + \frac{1}{\beta} D\big(f(w|Z^n)\|\pi(w)\big)\Big). && (0.176)
\end{aligned}$$

For any fixed dataset $z^n$, the posterior distribution $g(w)$ that minimizes the quantity inside the bracket is the solution to the optimization problem:

$$\begin{aligned}
&\min_{g(w)}\Big(\,\mathsf{E}_{W\sim g}\left(\mathcal{L}(W, z^n)\right) + \frac{1}{\beta} D(g\|\pi)\Big) \\
&\qquad = \frac{1}{\beta}\min_{g(w)}\Big(\,\mathsf{E}_{W\sim g}\left(\beta\mathcal{L}(W, z^n) - \ln \pi(W) + \ln g(W)\right)\Big) && (0.177)\\
&\qquad = \frac{1}{\beta}\min_{g(w)}\Big(D(g\|f_\beta) - \ln \mathsf{E}_{\tilde{W}\sim\pi}\left(e^{-\beta\mathcal{L}(\tilde{W}, z^n)}\right)\Big), && (0.178)
\end{aligned}$$

where $f_\beta(w|z^n)$ is the Gibbs algorithm in (0.158). Since $D(g\|f_\beta) \ge 0$, the minimum is achieved when $g(w) = f_\beta(w|z^n)$ and its value is

$$-\frac{1}{\beta}\ln \mathsf{E}_{\tilde{W}}\left(e^{-\beta\mathcal{L}(\tilde{W}, z^n)}\right)$$

By definition, the Gibbs algorithm uses this optimal posterior. Substituting this minimum value back into (0.176) and then into (0.173) completes the proof.

Under suitable conditions on the loss function, the first term on the right-hand side of (0.171) can be further bounded.

**Corollary 0.3.** Suppose that $\ell(w, Z)$ is $\sigma^2$-subgaussian for every $w \in \mathscr{W}$, and that the smoothness condition on $\ell$ in Problem 0.17 holds. Then the expected risk of the Gibbs algorithm (0.158) satisfies

$$\mathsf{E}(L(W)) \le \inf_{w \in \mathscr{W}} L(w) + \frac{d}{\beta} \ln \frac{A\beta}{d} + \frac{\beta\sigma^2}{2n} \tag{0.179}$$

for some constant $A > 0$ that depends on $\ell$.

**Proof.** From Problem 0.17, we have

$$-\frac{1}{\beta} \ln \mathsf{E}_{\tilde{W}} \Big( \exp(-\beta \mathcal{L}(\tilde{W}, z^n)) \Big) \le \inf_{w \in \mathscr{W}} \mathcal{L}(w, z^n) + \frac{d}{\beta} \ln \frac{A\beta}{d} \tag{0.180}$$

for some constant $A > 0$ that depends on $\ell$. Taking expectation with respect to $Z^n$ and using the fact that, for any collection of random variables $\{X_\alpha\}$, $\mathsf{E}(\min_\alpha \{X_\alpha\}) \le \min_\alpha \mathsf{E}(X_\alpha)$, we obtain

$$-\frac{1}{\beta} \mathsf{E}_{Z^n} \Big( \ln \mathsf{E}_{\tilde{W}} \big( e^{-\beta \mathcal{L}(\tilde{W}, Z^n)} \big) \Big) \le \mathsf{E} \big( \inf_{w \in \mathscr{W}} \mathcal{L}(w, Z^n) \big) + \frac{d}{\beta} \ln \frac{A\beta}{d} \tag{0.181}$$

$$\le \inf_{w \in \mathscr{W}} \mathsf{E} \big( \mathcal{L}(w, Z^n) \big) + \frac{d}{\beta} \ln \frac{A\beta}{d} \tag{0.182}$$

$$= \inf_{w \in \mathscr{W}} L(w) + \frac{d}{\beta} \ln \frac{A\beta}{d}. \tag{0.183}$$

Substituting in (0.171) completes the proof of the corollary.

Note that the bound in the corollary suggests choosing the inverse temperature $\beta = \sqrt{n}$, which yields excess risk scaling as $O\big(\max\{d, \sigma^2\}/\sqrt{n}\big)$, up to logarithmic factors.

## 0.6 PAC-BAYES BOUNDS

The Gibbs algorithm introduced in Section 0.5.2 is an example of a randomized learning algorithm, described by a conditional pdf $f(w|z^n)$. Its definition in (0.158) involves a pdf $\pi(w)$ that does not depend on the data $z^n$. This concept can be extended through the *PAC-Bayes* framework. Here, the term "Bayes" is used loosely, drawing an analogy to Bayes rule, which updates a data-independent prior into a data-dependent posterior.

The key insight behind PAC-Bayes bounds is that the generalization error of a randomized learning algorithm $f(w|z^n)$ can be controlled through the relative entropy $D(f(w|z^n) \| \pi(w))$ with respect to a chosen data-independent reference pdf $\pi(w)$. To illustrate the central ideas, we present one of the simplest such bounds.

We begin with the following consequence of the variational representation of relative entropy and the Markov inequality.

**Lemma 0.14.** Let $\pi(w)$ be a pdf supported on $\mathscr{W}$. Let $f(w|z^n)$ be a randomized learning algorithm. Let $\phi : \mathscr{W} \times \mathscr{Z}^n \to \mathbb{R}$ be a given function and assume that $\mathsf{E}_{\tilde{W}\sim\pi}\left(e^{\phi(\tilde{W},Z^n)}\right)$ is finite almost surely with respect to the data-distribution $f(z^n)$. Then, for any $\delta \in (0,1)$,

$$\mathsf{E}_{W\sim f(w|Z^n)}\left(\phi(W,Z^n)\right) \le \ln\left(\frac{1}{\delta}\,\mathsf{E}_{Z^n,\tilde{W}\sim\pi}\left(e^{\phi(\tilde{W},Z^n)}\right)\right) + D(f(w|Z^n)\|\pi(w)), \qquad (0.184)$$

where $\tilde{W} \sim \pi$ is independent of $Z^n$, holds with probability at least $1-\delta$ with respect to the distribution of $Z^n$.

**Proof.** Let $z^n$ be an outcome of the data $Z^n$. We use the variational representation of relative entropy in (0.139) with $p(w) = f(w|z^n)$, $q(w) = \pi(w)$, and $\psi(w) = \phi(w, z^n)$, to obtain

$$\mathsf{E}_{W\sim f(w|z^n)}\left(\phi(W,z^n)\right) \le \ln\left(\mathsf{E}_{\tilde{W}}\left(e^{\phi(\tilde{W},z^n)}\right)\right) + D(f(w|z^n)\|\pi(w)). \qquad (0.185)$$

This inequality holds for any fixed $z^n$ for which the right-hand side is well-defined, which is guaranteed by the assumption in the lemma. The expectation of $\mathsf{E}_{\tilde{W}\sim\pi}\left(e^{\phi(\tilde{W},Z^n)}\right)$ over the distribution of the data $Z^n$ is

$$\mathsf{E}_{Z^n}\left(\mathsf{E}_{\tilde{W}}\left(e^{\phi(\tilde{W},Z^n)}\right)\right) = \mathsf{E}_{Z^n,\tilde{W}}\left(e^{\phi(\tilde{W},Z^n)}\right),$$

since $\tilde{W}$ and $Z^n$ are independent. Applying the Markov inequality to $\mathsf{E}_{\tilde{W}}\left(e^{\phi(\tilde{W},Z^n)}\right)$, we obtain

$$\mathsf{Pr}_{Z^n}\left\{\mathsf{E}_{\tilde{W}}\left(e^{\phi(\tilde{W},Z^n)}\right) \le \frac{1}{\delta}\,\mathsf{E}_{Z^n,\tilde{W}}\left(e^{\phi(\tilde{W},Z^n)}\right)\right\} \ge 1-\delta.$$

Since the logarithm function is monotonically increasing, we can take the logarithm of the inequality on $\mathsf{E}_{\tilde{W}}\left(e^{\phi(\tilde{W},Z^n)}\right)$ to obtain

$$\mathsf{Pr}_{Z^n}\left\{\ln\left(\mathsf{E}_{\tilde{W}}\left(e^{\phi(\tilde{W},Z^n)}\right)\right) \le \ln\left(\frac{1}{\delta}\,\mathsf{E}_{Z^n,\tilde{W}}\left(e^{\phi(\tilde{W},Z^n)}\right)\right)\right\} \ge 1-\delta.$$

Combining with the inequality (0.185) completes the proof of the lemma.

Using this lemma, we obtain the following PAC-Bayes generalization bound.

**Theorem 0.10.** Let $\ell(w, Z)$ be $\sigma^2$-subgaussian for every $w \in \mathscr{W}$. Let $f(w|z^n)$ be a randomized learning algorithm and $\pi(w)$ be an arbitrary pdf supported on $\mathscr{W}$. Then, with probability at least $1-\delta$,

$$|L(W) - L_n(W)| \le \sqrt{\frac{2\sigma^2}{n-1}\Big(D(f(w|Z^n)\|\pi(w)) + \frac{1}{2}\ln n + \ln\frac{1}{\delta}\Big)}. \qquad (0.186)$$

**Proof.** We apply Lemma 0.14 with

$$\phi(w, z^n) = \frac{(n-1)}{2\sigma^2}\left(L(w) - \mathcal{L}(w, z^n)\right)^2. \qquad (0.187)$$

For each fixed $w \in \mathscr{W}$, the random variable

$$\sqrt{n/2\sigma^2}\big(L(w) - \mathcal{L}(w, Z^n)\big)$$

is (1/2)-subgaussian with zero mean. From Lemma 0.3 with $\lambda = (n-1)/n$, we obtain

$$\mathsf{E}\left(e^{\phi(w,Z^n)}\right) \le \sqrt{n} \quad \text{for every } w \in \mathscr{W}. \tag{0.188}$$

Consequently, for $\tilde{W} \sim \pi$ independent of $Z^n$,

$$\mathsf{E}_{Z^n,\tilde{W}}\left(e^{\phi(\tilde{W},Z^n)}\right) \le \sqrt{n}. \tag{0.189}$$

Applying Lemma 0.14 then gives, with probability at least $1-\delta$,

$$\mathsf{E}_{f(w|z^n)}\big(L(w) - \mathcal{L}(w, Z^n)\big) \le \frac{2\sigma^2}{n-1}\Big(\ln\frac{\sqrt{n}}{\delta} + D(f(w|Z^n)\|\pi(w))\Big). \tag{0.190}$$

Finally, the PAC-Bayes generalization bound (0.186) follows by applying the inequality $\mathsf{E}(|U|) \le \sqrt{\mathsf{E}(U^2)}$ to the random variable

$$U = |L(W) - \mathcal{L}(W, Z^n)| = |L(W) - L_n(W)| \quad \text{with } W \sim f(w|Z^n).$$

## 0.7 MINIMAX ESTIMATION FRAMEWORK

We now turn our attention to bounds on the performance of estimators. In Chapter 13 (Statistical Inference), we introduced the notion of dominance in estimation: an estimator $\hat{\theta}_1(X^n)$ of $\theta$ dominates another estimator $\hat{\theta}_2(X^n)$ if its mean squared error is no larger for every $\theta \in \Theta$. We also developed the Cramér–Rao bound, which provides a lower limit on the variance of unbiased estimators in regular parametric models. This classical result is inherently *local*—stated in terms of a single parameter value—and relies on restrictive assumptions, such as unbiasedness and differentiability. Consequently, it does not apply to many estimation problems of practical interest, particularly those involving biased estimators or nonparametric structure, such as probability density function estimation.

The minimax estimation approach, introduced in the seminal work of Wald (1950), provides a different perspective: instead of comparing estimators pairwise, it evaluates an estimator by its worst-case risk—the largest expected loss over all parameter values. Moreover, it seeks an estimator that minimizes this quantity. This *global* viewpoint is especially important in high-dimensional statistics, where the number of parameters may be comparable to or exceed the number of samples—a phenomenon often referred to as *the curse of dimensionality*; see, e.g., Tsybakov (2009). In such settings, certain parameter configurations—for example sparse vectors with many small entries or a few large but rare ones—can be intrinsically difficult to estimate reliably from limited data, and classical estimators such as the MLE may perform poorly in these challenging cases; see, e.g., Giraud (2021). Minimax analysis characterizes the fundamental performance limits of estimation by providing guarantees that hold uniformly over the entire probability model, in both parametric and nonparametric settings.

### 0.7.1 Definitions

Recall that in estimation, we are given a dataset and wish to infer a parameter (or a functional) of the unknown data distribution—for example, the mean, the covariance, or the distribution itself. To simplify the discussion, we assume that the unknown data distribution $p_{\text{data}}$ belongs to a parametric probability model $\mathscr{P} = \{p_\theta : \theta \in \Theta\}$, where every $p_\theta$ is a distribution on $\mathbb{R}^k$. As before, the data consists of i.i.d. samples $\mathbf{X}^n = (\mathbf{X}_1, \mathbf{X}_2, \dots, \mathbf{X}_n)$ drawn from $p_{\text{data}}$.

**Definition 0.16.** An *estimator* of $\theta$ is a function of the data $\hat{\theta} : \mathbb{R}^{kn} \to \Theta$, $\mathbf{X}^n \mapsto \hat{\theta}(\mathbf{X}^n)$.

While this framework extends readily to estimating functions or functionals of $\theta$, we focus on the estimation of $\theta$ itself for clarity of presentation.

**Definition 0.17.** The quality of an estimator $\hat{\theta}$ is evaluated using a *loss function* $\ell : \Theta \times \Theta \to \mathbb{R}^+$, $(\theta, \hat{\theta}) \mapsto \ell(\theta, \hat{\theta})$.

**Definition 0.18.** The *risk function* for the estimator $\hat{\theta}$ is defined as

$$R_n(\theta, \hat{\theta}) = \mathsf{E}_{\mathbf{X}^n \sim P_\theta^n}\left(\ell(\theta, \hat{\theta}(\mathbf{X}^n))\right), \tag{0.191}$$

where $P_\theta^n(\mathbf{x}^n) = \prod_{i=1}^n p_\theta(\mathbf{x}_i)$.

**Minimax estimation problem.** In the minimax framework, we compare the estimators based on their worst-case risk $\sup_{\theta\in\Theta} R_n(\theta, \hat{\theta})$, and the goal is to find the estimator that achieves the *minimax risk*

$$\mathscr{R}_n(\mathscr{P}, \ell) = \inf_{\hat{\theta}} \sup_{\theta\in\Theta} R_n(\theta, \hat{\theta}). \tag{0.192}$$

### 0.7.2 Examples

We consider examples where the minimax risk can be determined.

**Example 0.8** (Gaussian mean minimax estimator)**.** Let the dataset $X^n = (X_1, X_2, \dots, X_n)$, where $X_i \sim \mathrm{N}(\theta, \sigma^2)$ with known variance $\sigma^2$. Let $\hat{\theta}(X^n)$ denote an estimator of $\theta$. A natural loss function for this problem is the squared error loss $\ell(\theta, \hat{\theta}) = (\hat{\theta}(X^n) - \theta)^2$. The minimax risk is therefore

$$\mathscr{R}_n(\mathscr{P}, \ell) = \inf_{\hat{\theta}} \sup_{\theta} \mathsf{E}_{X^n}\left((\hat{\theta}(X^n) - \theta)^2\right).$$

We show that the sample mean $\bar{X} = \frac{1}{n}\sum_{i=1}^n X_i$ attains the minimax risk. First note that

$$\mathsf{E}_{X^n}\left((\bar{X} - \theta)^2\right) = \frac{\sigma^2}{n},$$

independent of $\theta$. Hence,

$$\mathscr{R}_n(\mathscr{P}, \ell) \le \sup_{\theta} R(\theta, \bar{X}) \le \frac{\sigma^2}{n}.$$

To establish a matching lower bound, we use a Bayesian argument. Let $\mathbf{\Theta} \sim \pi(\theta)$. Replacing $\sup_\theta$ in the minimax risk by expectation over $\mathbf{\Theta}$ yields

$$\mathscr{R}_n(\mathscr{P}, \ell) \geq \min_{\hat{\theta}} \mathsf{E}_\pi \left( R_n(\mathbf{\Theta}, \hat{\theta}) \right). \tag{0.193}$$

The right-hand side corresponds to the minimum MSE between $\mathbf{\Theta}$ and its estimate $\hat{\theta}(X^n)$, which is attained by $\hat{\theta}^*_\pi(X^n) = \mathsf{E}(\mathbf{\Theta}|X^n)$. The corresponding minimum MSE is $\mathsf{E}_{X^n}\left( \operatorname{Var}(\mathbf{\Theta}|X^n) \right)$. Now choosing $\mathbf{\Theta} \sim \mathrm{N}(0, \gamma^2)$ for $\gamma^2 > 0$, we have

$$\hat{\theta}^*_\pi(X^n) = \mathsf{E}(\mathbf{\Theta}|X^n) = \frac{n\gamma^2}{n\gamma^2 + \sigma^2}\bar{X}, \tag{0.194}$$

$$\min_{\hat{\theta}} \mathsf{E}_\pi \left( R_n(\mathbf{\Theta}, \hat{\theta}(X^n)) \right) = \mathsf{E}_{X^n}\left( \operatorname{Var}(\mathbf{\Theta}|X^n) \right) = \frac{\sigma^2}{n + (\sigma/\gamma)^2}. \tag{0.195}$$

Hence,

$$\mathscr{R}_n(\mathscr{P}, \ell) \geq \lim_{\gamma^2 \to \infty} \mathsf{E}_{X^n}\left( \operatorname{Var}(\mathbf{\Theta}|X^n) \right) = \frac{\sigma^2}{n}. \tag{0.196}$$

Since the sample mean $\bar{X}$ is an estimator whose maximum risk achieves this lower bound, it is a minimax estimator.

**Example 0.9** (Hypothesis testing)**.** Even though in Chapter 13 we introduced hypothesis testing as a problem distinct from estimation, it can be formulated within the minimax estimation framework. The probability family for the two-hypothesis testing problem is $\mathscr{P} = \{p_\theta : \theta \in \{0, 1\}\}$. We use the 0-1 loss function, which penalizes the incorrect decision:

$$\ell(\theta, \hat{\theta}) = \begin{cases} 1 & \text{if } \hat{\theta} \neq \theta, \\ 0 & \text{otherwise.} \end{cases}$$

The risk of a test $\hat{\theta}(X^n) \in \{0, 1\}$ is its probability of error for $X^n \sim P^n_\theta$ is

$$R_n(\theta, \hat{\theta}) = \mathsf{Pr}\{\hat{\theta}(X^n) \neq \theta\}. \tag{0.197}$$

This risk function corresponds to the two standard probabilities of error $\alpha_n = R_n(0, \hat{\theta}) = P^n_0(\{\hat{\theta}(X^n) = 1\})$ and $\beta_n = R_n(1, \hat{\theta}) = P^n_1(\{\hat{\theta}(X^n) = 0\})$. The minimax risk is the smallest achievable value of the worst-case error probability

$$\mathscr{R}_n(\mathscr{P}, \ell) = \min_{\hat{\theta}} \max_{\theta \in \{0,1\}} R_n(\theta, \hat{\theta}) = \min_{\hat{\theta}} \max\{\alpha_n, \beta_n\}.$$

The estimator that achieves this minimum is the likelihood ratio test:

$$\hat{\theta}^*(x^n) = \begin{cases} 0 & \text{if } \dfrac{P_0(x^n)}{P_1(x^n)} > t, \\ 1 & \text{otherwise,} \end{cases}$$

where $t$ is chosen so that $\alpha_n = \beta_n$ (For simplicity we assume that such a test exists.)

To see why this test is optimal, we invoke the Neyman–Pearson lemma (Chapter 13). Let $\hat{\theta}^*$ be the likelihood ratio test, chosen such that its error probabilities are equal: $\alpha_n^* = \beta_n^*$. The maximum risk for this test is therefore $\alpha_n^* = \beta_n^* = R^*$.

Now, consider any other test $\hat{\theta}$ with error probabilities $\alpha_n$ and $\beta_n$. To show that its maximum risk cannot be less than $R^*$, we consider two possible cases. If $\alpha_n \geq \alpha_n^*$, then the maximum risk for the test $\hat{\theta}$ is $\max\{\alpha_n, \beta_n\} \geq \alpha_n$. Since $\alpha_n \geq \alpha_n^* = R^*$, the maximum risk of $\hat{\theta}$ is at least $R^*$. On the other hand, if $\alpha_n < \alpha_n^*$, the Neyman–Pearson lemma states that for a given $\alpha_n$, the likelihood ratio test achieves the minimum possible $\beta_n$. Since the test $\hat{\theta}$ has $\alpha_n < \alpha_n^*$, it must necessarily have $\beta_n > \beta_n^*$. The maximum risk for $\hat{\theta}$ is then $\max\{\alpha_n, \beta_n\} \geq \beta_n$. Since $\beta_n > \beta_n^* = R^*$, the maximum risk of $\hat{\theta}$ is strictly greater than $R^*$.

In both cases, the maximum risk of any other test $\hat{\theta}$ is greater than or equal to $R^*$, the maximum risk of $\hat{\theta}^*$. Thus, $\hat{\theta}^*$ minimizes the maximum possible risk and is the minimax optimal test.

We next recall the minimax probability estimation problem discussed in Chapter 12 (Universal Compression).

**Example 0.10** (Minimax sequential probability estimation). Assume the data $X_1, X_2, \ldots$ are i.i.d. drawn from $p_\theta \in \mathscr{P}(\mathcal{X})$, where $\mathcal{X}$ is a finite set. A sequential estimator $q(x|X^n)$ maps the observed sequence of length $n$ to a probability distribution on $\mathcal{X}$. The performance of the estimator is measured by the *relative entropy loss*

$$\ell(p_\theta, q(X^n)) = \sum_{x \in \mathcal{X}} p_\theta(x) \log \frac{p_\theta(x)}{q(x|X^n)} \tag{0.198}$$

$$= D(p_\theta(x) \| q(x|X^n)). \tag{0.199}$$

The risk is the expected loss

$$R_n(p_\theta, q) = \mathsf{E}_{X^n \sim p_\theta^n}\left(D\left(p_\theta(x) \| q(x|X^n)\right)\right)$$

The goal is to find minimax risk

$$\mathscr{R}_n(\mathscr{P}, \ell) = \inf_q \sup_{p_\theta \in \mathscr{P}} R_n(p_\theta, q).$$

As discussed in Chapter 12, for a binary alphabet, the minimax risk $\mathscr{R}_n(\mathscr{P}, \ell) = \frac{1}{2n} \cdot (1 + o(1))$. This optimal risk is asymptotically achieved by the Braess–Sauer estimator.

## 0.8 LOWER BOUNDS ON MINIMAX RISK

We have discussed examples in which the minimax risk can be evaluated exactly or asymptotically. In general, however, computing the minimax risk is difficult. Instead, we seek computable lower bounds that allow us to assess the performance of any estimator and to characterize how the minimax risk scales with the number of samples.

For developing lower bounds, we assume that the parameter space $\Theta$ is equipped with a metric (or pseudometric) $\rho$. We further assume that the loss function can be written as

$$\ell(\theta, \hat{\theta}) = \Phi\big(\rho(\theta, \hat{\theta}(\mathbf{X}^n))\big),$$

where $\Phi : \mathbb{R}^+ \to \mathbb{R}^+$ is nondecreasing with $\Phi(0) = 0$. For example, in the Gaussian mean estimation problem, we may take $\rho(\theta, \hat{\theta}(X^n)) = |\hat{\theta}(X^n) - \theta|$ and $\Phi(a) = a^2$, which yields the squared error loss $\ell(\theta, \hat{\theta}(X^n)) = (\hat{\theta}(X^n) - \theta)^2$.

The approach to deriving these lower bounds, as detailed in the next sections, proceeds in three steps:

1. We construct an appropriate packing of the parameter space $\Theta$ and use it to relate the minimax risk to the probability of error in a corresponding multi-hypothesis testing problem.
2. We apply Fano's inequality to obtain a lower bound on the probability of error in terms of the mutual information between a randomly selected hypothesis and the data.
3. Because the mutual information can be difficult to evaluate directly, we introduce two methods for bounding it and illustrate them via examples.

### 0.8.1 From estimation to hypothesis testing

Suppose we have a finite set $\{\theta_1, \theta_2, \dots, \theta_m\} \subseteq \Theta$ such that $\rho(\theta_j, \theta_{j'}) \ge 2\delta$ for every $j \ne j'$ (we will refer to this set as $2\delta$-packing of $\Theta$ even though the $>$ in the definition of a packing is replaced with $\ge$).

Now, consider the following $m$-ary hypothesis testing problem: we let $J \sim \mathrm{Unif}\{1, 2, \dots, m\}$ and conditioned on $J = j$, we draw i.i.d. samples $\mathbf{X}_1, \mathbf{X}_2, \dots, \mathbf{X}_n$ from $p_{\theta_j}$. A testing function $\psi : \mathbb{R}^{kn} \to \{1, 2, \dots, m\}$ is then used to estimate $J$.

For $j = 1, 2, \dots, m$, we write $p_{\theta_j}(\mathbf{x}) \in \mathscr{P}$ as $p_j$, and let $P_j$ be the corresponding probability measure, that is, $P_j(\mathcal{A}) = \sum_{\mathbf{x} \in \mathcal{A}} p_j(\mathbf{x})$ for any measurable set $\mathcal{A} \subseteq \mathbb{R}^k$. The probability of error for $\psi$ is then given by

$$\Pr\{\psi(\mathbf{X}^n) \ne J\} = \frac{1}{m} \sum_{j=1}^{m} P_j^n(\{\psi(\mathbf{X}^n) \ne j\}), \tag{0.200}$$

where $P_j^n$ is the product measure corresponding to $P_j$. We now bound the minimax risk in (0.192) in terms of the minimum probability of error over $\psi$.

**Theorem 0.11.** Let $\mathscr{P} = \{p_\theta : \theta \in \Theta\}$ be a probability model, and let $\ell$ be a loss function. The minimax risk of estimating $\theta$ based on $n$ i.i.d. samples $\mathbf{X}_1, \mathbf{X}_2, \dots, \mathbf{X}_n$ is lower bounded as

$$\mathscr{R}_n(\mathscr{P}, \ell) \ge \Phi(\delta) \inf_{\psi} \Pr\{\psi(\mathbf{X}^n) \ne J\}, \tag{0.201}$$

where the infimum is over all measurable test functions $\psi$ and $J$ indexes a randomly chosen parameter from a suitable $2\delta$-packing of $\Theta$.

There is an inherent tension in the choice $\delta$ in this bound. Because $\Phi$ is nondecreasing, the first term in the bound becomes looser as $\delta$ decreases. However, decreasing $\delta$ enlarges the size of the $2\delta$-packing, hence increasing the number of hypotheses and making the testing problem harder. As a result, the probability of error term typically grows as $\delta$ increases. Thus the selection of $\delta$ needs to trade off the parameter space approximation against the testing problem complexity.

**Proof.** For simplicity, we consider the case of a single sample $n = 1$ and drop the subscript $n$ from $R_n$ and $\mathscr{R}_n$. The extension to $n > 1$ is straightforward. For any $\theta \in \Theta$ and any estimator $\hat{\theta}$, we have

$$R(\theta, \hat{\theta}) \geq \Phi(\delta) \operatorname{Pr}\{\ell(\theta, \hat{\theta})) \geq \Phi(\delta)\} \tag{0.202}$$

$$\geq \Phi(\delta) \operatorname{Pr}\{\rho(\theta, \hat{\theta}) \geq \delta\}, \tag{0.203}$$

where (0.202) follows from the Markov inequality, and the last step follows from the monotonicity of $\Phi$. Thus,

$$\sup_\theta R(\theta, \hat{\theta}) \geq \Phi(\delta) \sup_\theta \operatorname{Pr}\{\rho(\theta, \hat{\theta}) \geq \delta\}. \tag{0.204}$$

Now, for every $j = 1, 2, \ldots, m$, $\sup_\theta \operatorname{Pr}\{\rho(\theta, \hat{\theta}) \geq \delta\} \geq P_j(\{\rho(\theta_j, \hat{\theta}) \geq \delta\})$. Hence,

$$\sup_\theta \operatorname{Pr}\{\rho(\theta, \hat{\theta}) \geq \delta\} \geq \frac{1}{m} \sum_{j=1}^{m} P_j(\{\rho(\theta_j, \hat{\theta}) \geq \delta\}). \tag{0.205}$$

We design a test for the hypothesis testing problem using the original estimator $\hat{\theta}$. Specifically, we define the "nearest-neighbor" test

$$\psi(\mathbf{X}) = \operatorname*{arg\,min}_{j \in \{1,2,\ldots,m\}} \rho(\theta_j, \hat{\theta}(\mathbf{X})), \tag{0.206}$$

with ties broken arbitrarily. We claim that if $\theta_j$ is the true parameter, then the event $\{\rho(\theta_j, \hat{\theta}) < \delta\}$ guarantees that the test $\psi$ yields the correct outcome. To see this, note that for any other index $j' \neq j$,

$$\rho(\theta_{j'}, \hat{\theta}) \geq \rho(\theta_j, \theta_{j'}) - \rho(\theta_j, \hat{\theta}) \tag{0.207}$$

$$> 2\delta - \delta \tag{0.208}$$

$$= \delta, \tag{0.209}$$

where (0.207) follows from the triangle inequality and (0.209) follows since by design $\rho(\theta_j, \theta_{j'}) \geq 2\delta$ and by the definition of the test, $\rho(\theta_j, \hat{\theta}) < \delta$. Hence, for $j = 1, 2, \ldots, m$, $\rho(\theta_j, \hat{\theta}) < \delta$ implies that $\psi(\mathbf{X}) = j$, and $P_j(\{\rho(\theta_j, \hat{\theta}(\mathbf{X})) \geq \delta\}) \geq P_j(\{\psi(\mathbf{X}) \neq j\})$. Taking the average on both sides, we have shown that

$$\frac{1}{m}\sum_{j=1}^{m} P_j(\{\rho(\theta_j, \hat{\theta}) \geq \delta\}) \geq \Pr\{\psi(\mathbf{X}) \neq J\}. \tag{0.210}$$

Combining this inequality with (0.205) and (0.204) and taking the infimum over $\hat{\theta}$ on both sides completes the proof of the theorem.

### 0.8.2 Applying Fano's inequality

We now focus on the hypothesis testing problem defined in the previous section. Recall Fano's inequality, which was introduced in Chapter 2 (Entropy, Relative Entropy, and Mutual Information) and has since been applied numerous times to establish converses of coding theorems. We will consider the following specialized version of this inequality.

Let $J \sim \text{Unif}\{1, 2, \ldots, m\}$ and let $\mathbf{X}^n$ be an observation taking values in $\mathbb{R}^{kn}$, which can be viewed as an output of a channel $p(\mathbf{x}^n|j) = \prod_{i=1}^{n} p_j(\mathbf{x}_i)$ whose input is $J$. Finally, let $\psi : \mathbb{R}^{kn} \to \{1, 2, \ldots, m\}$ be an estimate of $J$ based on $\mathbf{X}^n$ and define the probability of error $P_e = \Pr\{\psi(\mathbf{X}^n) \neq J\}$, then

$$H(J|\psi(\mathbf{X}^n)) \leq H(J|\mathbf{X}^n) \tag{0.211}$$
$$\leq \mathsf{H}(P_e) + P_e \log(m-1) \tag{0.212}$$
$$\leq 1 + P_e \log m, \tag{0.213}$$

where the second step follows from Fano's inequality. Now, using the fact that $J$ is uniformly distributed, we have

$$I(J; \mathbf{X}^n) \geq \log m - 1 - P_e \log m \tag{0.214}$$
$$= (1 - P_e)\log m - 1. \tag{0.215}$$

Rearranging, and taking the infimum over all estimators $\psi$ we obtain the following bound.

**Lemma 0.15.** Let $J \sim \text{Unif}\{1, 2, \ldots, m\}$ and $\mathbf{X}^n|\{J = j\} \sim P_j^n$. Then,

$$\inf \Pr\{\psi(\mathbf{X}^n) \neq J\} \geq 1 - \frac{I(J; \mathbf{X}^n) + 1}{\log m}, \tag{0.216}$$

where the infimum is over all measurable test functions $\psi$.

Combining this lemma with Theorem 0.11 gives the following lower bound on the minimax risk.

**Theorem 0.12.** Let $\{\theta_1, \theta_2, \ldots, \theta_m\}$ be any $2\delta$-packing of $\Theta$, and let $J \sim \text{Unif}\{1, 2, \ldots, m\}$ and $\mathbf{X}^n|\{J = j\} \sim P_j^n$. Then, the minimax risk is lower bounded as

$$\mathscr{R}_n(\mathscr{P}, \ell) \geq \Phi(\delta)\Big(1 - \frac{I(J; \mathbf{X}^n) + 1}{\log m}\Big). \tag{0.217}$$

The mutual information $I(\mathbf{X}^n; J)$ appearing in Theorem 0.12 is generally difficult to compute and does not directly reveal how the minimax risk scales with the sample size $n$. To address this, we develop two techniques for upper bounding this mutual information, yielding more tractable and interpretable lower bounds on the minimax risk.

## 0.9 THE LOCAL FANO METHOD

The *local Fano method* bounds the mutual information by considering a carefully chosen finite subset of parameters that are well separated under the metric $\rho$. Intuitively, if the distributions associated with these parameters are difficult to distinguish from $n$ samples, then any estimator must incur a nontrivial error. The key step in this approach is the following upper bound on the mutual information.

**Lemma 0.16.** Let $J \sim \mathrm{Unif}\{1, 2, \dots, m\}$ and $\mathbf{X}^n | \{J = j\} \sim P_j^n$. Then, the mutual information between $J$ and $\mathbf{X}^n$ is bounded as

$$I(J; \mathbf{X}^n) \le \frac{n}{m^2} \sum_{j=1}^{m} \sum_{j'=1}^{m} D(p_j \| p_{j'}). \tag{0.218}$$

**Proof.** First we can express the mutual information in terms of divergence as

$$\begin{aligned} I(J; \mathbf{X}^n) &= \frac{1}{m} \sum_{j=1}^{m} D(p_j(\mathbf{x}^n) \| p(\mathbf{x}^n)) && (0.219)\\ &= \frac{1}{m} \sum_{j=1}^{m} D\Big(p_j(\mathbf{x}^n) \Big\| \frac{1}{m} \sum_{j'=1}^{m} p_{j'}(\mathbf{x}^n)\Big) && (0.220)\\ &\le \frac{1}{m^2} \sum_{j=1}^{m} \sum_{k=1}^{m} D(P_j^n \| P_{j'}^n) && (0.221)\\ &= \frac{n}{m^2} \sum_{j=1}^{m} \sum_{j'=1}^{m} D(p_j \| p_{j'}), && (0.222) \end{aligned}$$

where step (0.221) follows by the convexity of relative entropy, and the last step follows from the fact that $\mathbf{X}_1, \mathbf{X}_2, \dots, \mathbf{X}_n$ are i.i.d., hence $p_j(\mathbf{x}^n) = \prod_{i=1}^{n} p_j(\mathbf{x}_i)$.

Combining this lemma with Theorem 0.12, we arrive at the following bound on the minimax risk.

**Theorem 0.13.** Let $\{\theta_1, \theta_2, \dots, \theta_m\}$ be any $2\delta$-packing of $\Theta$, and let $J \sim \mathrm{Unif}\{1, 2, \dots, m\}$ and $\mathbf{X}^n | \{J = j\} \sim P_j^n$. Then, the minimax risk is lower bounded as

$$\mathscr{R}_n(\mathscr{P}, \ell) \ge \Phi(\delta)\Big(1 - \frac{(n/m^2) \sum_{j=1}^{m} \sum_{j'=1}^{m} D(p_j \| p_{j'}) + 1}{\log m}\Big). \tag{0.223}$$

This theorem motivates a method known as *local Fano* for deriving useful minimax bounds. The idea is to construct a $2\delta$-packing $\{\theta_1, \theta_2, \ldots, \theta_m\}$ of a small neighborhood in $\Theta$, so that for all $j, j'$,

$$D(p_{\theta_j} \| p_{\theta_{j'}}) \leq c\delta^2, \tag{0.224}$$

for some constant $c > 0$, that is, the distributions corresponding to the points in the packing are close in relative entropy. We then choose $\delta$ sufficiently small to optimize the bound of Theorem 0.13. This is illustrated in the following.

**Example 0.11** (Estimating the mean of a Gaussian)**.** Let $\mathscr{P} = \{\mathrm{N}(\mathbf{x}; \theta, \sigma^2 I),\ \theta \in \mathbb{R}^k\}$ be a family of $k$-variate Gaussian distributions. Let $\rho(\theta, \theta') = \|\theta - \theta'\|_2$ and $\Phi(a) = a^2$. Consider the Euclidean ball centered on $\theta = \mathbf{0}$ with radius $r = 4\delta$. From Example 0.3 and Lemma 0.6, we have

$$\log M(2\delta; B(4\delta), \rho) \geq \log N(2\delta; B(4\delta), \rho) \geq k \log\left(\frac{4\delta}{2\delta}\right) = k \log 2. \tag{0.225}$$

Hence, there exists a $2\delta$-packing $\{\theta_1, \theta_2, \ldots, \theta_m\}$ with $\log m \geq k$. Then for this packing, we have

$$\|\theta_j\|_2 \leq 4\delta, \quad j \in \{1, 2, \ldots, m\} \tag{0.226}$$

$$2\delta \leq \|\theta_j - \theta_{j'}\|_2 \leq 8\delta, \quad j, j' \in \{1, 2, \ldots, m\},\ j \neq j', \tag{0.227}$$

where the upper bound follows by the triangle inequality.

From the formula for the relative entropy of two Gaussian distributions, for every $j, j' \in \{1, 2, \ldots, m\}$,

$$D(P_j \| P_{j'}) = \frac{\|\theta_j - \theta_{j'}\|_2^2}{2\sigma^2} \log e \tag{0.228}$$

$$\leq \frac{32\delta^2 \log e}{\sigma^2}, \tag{0.229}$$

where the last step follows from the upper bound in (0.227). Substituting in the bound of Theorem 0.13, we obtain

$$\mathscr{R}_n(\mathscr{P}, \ell) \geq \delta^2\Big(1 - \frac{32n\delta^2 \log e/\sigma^2 + 1}{k}\Big). \tag{0.230}$$

Now, choosing $\delta^2 = \sigma^2 k/(64n \log e)$ gives $(32n\delta^2 \log e/\sigma^2) + 1) = (k/2 + 1)$. Substituting in (0.230), we have shown that for $k \geq 3$,

$$\mathscr{R}_n(\mathscr{P}, \ell) \geq \delta^2\Big(1 - \frac{k/2 + 1}{k}\Big) \tag{0.231}$$

$$\geq \frac{\sigma^2 k}{64n \log e}(1/2 - 1/3) \tag{0.232}$$

$$= \frac{\sigma^2 k}{384n \log e}. \tag{0.233}$$

This shows that no estimator of the Gaussian mean can have risk that decays faster than order $k/n$. Since the risk of the sample mean $\hat{\theta} = \frac{1}{n}\sum_{i=1}^{n}\mathbf{X}_i$ decays as $\sigma^2 k/n$, it is order optimal.

The following example demonstrates how minimax bounds can be applied in nonparametric estimation.

**Example 0.12** (Density estimation)**.** (Wainwright 2019) Consider the family of probability density functions:

$$\mathscr{F} = \Big\{f : [0,1] \to [c_l, c_u], |f''(t)| \le c_d \text{ for every } t \in [0,1], \int_0^1 f(t)\,dt = 1\Big\}, \quad (0.234)$$

where $c_l < 1 < c_u$ and $c_d > 1$ are positive real numbers. We measure the estimation error using the Hellinger distance (Problem **??**), defined as

$$D_{\mathrm{H}}(f\|f') = \frac{1}{\sqrt{2}}\Big(\int_0^1 \Big(\sqrt{f(t)} - \sqrt{f'(t)}\Big)^2\,dt\Big)^{1/2}. \quad (0.235)$$

Taking $\Phi(a) = a^2$, the corresponding loss function is $\ell(f, f') = D_{\mathrm{H}}^2(f\|f')$.

We bound the minimax risk using the local Fano method by constructing a packing in a neighborhood of the uniform pdf $f(t) = \mathrm{Unif}[t; 0, 1]$.

Let $\phi : [0,1] \to \big[-1/2, 1/2\big]$ be an arbitrary twice-differentiable function with

$$\int_0^1 \phi(t)dt = 0, \quad \int_0^1 (\phi(t))^2 dt < \infty, \quad \int_0^1 (\phi'(t))^2 dt < \infty. \quad (0.236)$$

Let $\phi_j$, $j = 1, 2, \ldots, k$, be $\phi/k^2$ rescaled to be supported in the interval $[(j-1)/k, j/k]$, i.e.,

$$\phi_j(t) = \frac{1}{k^2}\,\phi\big(kt - (j-1)\big)\mathbb{1}_{[(j-1)/k, j/k]}(t). \quad (0.237)$$

From Problem 0.13, there exists a $(k/4)$-packing $\mathcal{S}$ of the cube $\{-1, +1\}^k$ in Hamming distance of size at least

$$\log m = kD\big(\mathrm{Bern}(1/4)\|\mathrm{Bern}(1/2)\big) \ge k/10.$$

We claim that the set $\big\{f_{\mathbf{b}}(t) = 1 + C_1\sum_{j=1}^{k} b_j\phi_j(t),\ \mathbf{b} \in \mathcal{S}\big\}$ is a $\delta$-packing of $\mathscr{F}$ in Hellinger distance with

$$\delta^2 = \frac{C_1^2}{8k^4}\int_0^1 (\phi(t))^2\,dt, \quad (0.238)$$

where $C_1$ is a sufficiently small constant (that depends only on $\phi$) such that $f_{\mathbf{b}} \in \mathscr{F}$ and $f_{\mathbf{b}}(t) \ge 1/2$ for every $t \in [0,1]$ and $\mathbf{b} \in \{-1,+1\}^n$. That such $C_1$ exists can be seen from the fact that $f_{\mathbf{b}} \to 1$ and $f_{\mathbf{b}} \to 0$ uniformly in $\mathbf{b}$ and $t$ as $C_1 \to 0$.

To show this, let $\mathbf{b} \neq \mathbf{b}'$ be elements of the packing $\mathcal{S}$. Let $\mathcal{I} = \{j : b_j \neq b_j'\}$. By the assumption that $\mathcal{S}$ is a $(k/4)$-packing, $|\mathcal{I}| > k/4$.

$$
\begin{aligned}
D_{\mathrm{H}}^2(f_{\mathbf{b}}, f_{\mathbf{b}'}) &= \frac{1}{2}\int_0^1 \left(\sqrt{f_{\mathbf{b}}(x)} - \sqrt{f_{\mathbf{b}'}(x)}\right)^2 dx && (0.239)\\
&\geq \frac{1}{2}\sum_{j\in\mathcal{I}}\int_{(j-1)/k}^{j/k} \left(\sqrt{f_{\mathbf{b}}(x)} - \sqrt{f_{\mathbf{b}'}(x)}\right)^2 dx && (0.240)\\
&= \frac{1}{2}\sum_{j\in\mathcal{I}}\int_{(j-1)/k}^{j/k} \frac{\left(f_{\mathbf{b}}(x) - f_{\mathbf{b}'}(x)\right)^2}{\left(\sqrt{f_{\mathbf{b}}(x)} + \sqrt{f_{\mathbf{b}'}(x)}\right)^2} dx && (0.241)\\
&\geq \frac{1}{2}\sum_{j\in\mathcal{I}}\int_{(j-1)/k}^{j/k} \frac{\left(f_{\mathbf{b}}(x) - f_{\mathbf{b}'}(x)\right)^2}{4} dx && (0.242)\\
&= \frac{1}{2}C_1^2\sum_{j\in\mathcal{I}}\int_{(j-1)/k}^{j/k} (\phi_j(x))^2\, dx && (0.243)\\
&= C_1^2 \cdot \frac{k}{8k^5}\int_0^1 (\phi(x))^2\, dx. && (0.244)
\end{aligned}
$$

It remains to bound the pairwise relative entropy. Again let $\mathbf{b} \neq \mathbf{b}'$ be elements of the packing $S$. By construction, we are guaranteed that $f_{\mathbf{b}}, f_{\mathbf{b}'}$ are bounded below by 1/2. Then, from Problem 0.19, we have

$$
\begin{aligned}
D(f_{\mathbf{b}}\|f_{\mathbf{b}'}) &\leq \frac{4}{\ln 2}D_{\mathrm{H}}^2(f_{\mathbf{b}}, f_{\mathbf{b}'}) && (0.245)\\
&\leq \frac{4}{k^4\ln 2}\int_0^1 \left(\phi(x)\right)^2 dx. && (0.246)
\end{aligned}
$$

Hence,

$$
(n/m^2)\sum_{\mathbf{b}\in\mathcal{S}}\sum_{\mathbf{b}'\in\mathcal{S}} D(f_{\mathbf{b}}\|f_{\mathbf{b}'}) \leq 16n\delta^2. \tag{0.247}
$$

Now, we choose $k = C_2 n^{1/5}$ for sufficiently large constant $C_2$ (depending on $\phi$ and the choice of $C_1$) such that

$$
\begin{aligned}
\frac{(n/m^2)\sum_{\mathbf{b}\in S}\sum_{\mathbf{b}'\in S} D(f_{\mathbf{b}}\|f_{\mathbf{b}'}) + 1}{\log m} &\leq \frac{16n\delta^2 + 1}{k/10} && (0.248)\\
&\leq 170C_1^2\frac{n}{k^5}\int_0^1 (\phi(x))^2 dx && (0.249)\\
&\leq \frac{1}{2}. && (0.250)
\end{aligned}
$$

Applying Theorem 0.13, we obtain the bound

$$
\mathscr{R}_n(\mathscr{P}, \ell) \geq \frac{1}{2}\Phi(\delta) \tag{0.251}
$$

$$= \frac{1}{2}\delta^2 \tag{0.252}$$

$$= C \cdot n^{-4/5}, \tag{0.253}$$

where

$$C = \frac{C_1^2}{16C_2^4}\int_0^1 \left(\phi(t)\right)^2 dt$$

is a constant that does not depend on the number of samples $n$.

We note that there exist nonparametric density estimators that achieve risk of order $n^{-4/5}$ in this setting (Tsybakov 2009). Consequently, the above lower bound is order-optimal.

## 0.10 THE GLOBAL FANO METHOD

In the *global Fano method*, the number of hypotheses is chosen to be the packing number $M(2\delta; \Theta, \rho)$ of the entire parameter space $\Theta$ (or a suitable lower bound). The mutual information term in (0.217) is then upper bounded using the covering number for the set of distributions $\mathscr{P}$ with respect to relative entropy.

**Definition 0.19.** The set $\{q_1, q_2, \ldots, q_K\}$ forms a *relative entropy $\epsilon$-net* of $\mathscr{P}$ if for every $p \in \mathscr{P}$, there exists some $j$, such that $D(p\|q_j) \le \epsilon^2$. The *relative entropy $\epsilon$-covering number* $K(\epsilon; \mathscr{P})$ is the minimum $K$, such that there exists a relative entropy $\epsilon$-net of $\mathscr{P}$.

We can now establish the following bound on the minimax risk.

**Theorem 0.14.** Let $\mathscr{P} = \{p_\theta : \theta \in \Theta\}$ be a probability model. For any $\epsilon, \delta > 0$, the minimax risk of estimating $\theta$ is lower bounded as

$$\mathscr{R}_n(\mathscr{P}, \ell) \ge \Phi(\delta)\Big(1 - \frac{\log K(\epsilon; \mathscr{P}) + n\epsilon^2 + 1}{\log M(2\delta; \Theta, \rho)}\Big), \tag{0.254}$$

where $K(\epsilon; \mathscr{P})$ is the relative entropy $\epsilon$-covering number of the model class $\mathscr{P}$, and $M(2\delta; \Theta, \rho)$ is the packing number of the parameter space.

**Proof.** Consider the bound in Theorem 0.12. Let the number of hypotheses $m$ be the packing number at resolution $2\delta$, that is, $m = M(2\delta; \Theta, \rho)$.

We upper bound the mutual information term as follows. Denote by $\{p_1, p_2, \ldots, p_m\}$ the distributions corresponding to the $m$ hypotheses, and define their average distribution by $\bar{p} = (1/m)\sum_{j=1}^m p_j$. Then, for any $q \in \mathscr{P}$, we have

$$I(J; \mathbf{X}^n) = \sum_j \frac{1}{m} D(p_j\|\bar{p}) \tag{0.255}$$

$$= \sum_j \frac{1}{m} D(p_j\|q) + \mathsf{E}_{\bar{p}}\left(\log\frac{q}{\bar{p}}\right) \tag{0.256}$$

$$= \sum_j \frac{1}{m} D(p_j \| q) - D(\bar{p} \| q) \tag{0.257}$$

$$\le \sum_j \frac{1}{m} D(p_j \| q), \tag{0.258}$$

where the last step follows from the nonnegativity of relative entropy.

Now, let $K = K(\epsilon; \mathscr{P})$ and let $\{q_1, q_2, \dots, q_K\}$ be a relative entropy $\epsilon$-net of $\mathscr{P}$, that is, for every $p \in \mathscr{P}$, $\min_i D(p\|q_i) \le \epsilon^2$. Define $\bar{q} = (1/K)\sum_j q_j$, and for each $j$, let $q_{i(j)}$ be the distribution attaining $\min_i D(p_j\|q_i)$, so that $D(p_j\|q_{i(j)}) \le \epsilon^2$. Then, for any $j = 1, 2, \dots, m$,

$$D(p_j \| \bar{q}) = \mathsf{E}_{p_j}\Big( \log \frac{p_j}{(1/K)\sum_{i=1}^K q_i} \Big) \tag{0.259}$$

$$\le \mathsf{E}_{p_j}\Big( \log \frac{p_j}{(1/K) q_{i(j)}} \Big) \tag{0.260}$$

$$= \log K + D(p_j \| q_{i(j)}) \tag{0.261}$$

$$\le \log K + \epsilon^2. \tag{0.262}$$

For i.i.d. samples $\mathbf{X}_1, \mathbf{X}_2, \dots, \mathbf{X}_n$ drawn from $p_j$, we replace $p_j$ with $P_j^n$, $q_j$ with $Q_j^n$, and $q$ with $(1/K)\sum_i Q_i^n$ in the above derivation. This yields the bound

$$I(J; \mathbf{X}^n) \le \log K + n\epsilon^2. \tag{0.263}$$

Substituting this bound in (0.217) establishes the theorem.

The following example demonstrates the use of the above bound and shows how the parameters $\epsilon$ and $\delta$ can be judiciously chosen to optimize it.

**Example 0.13** (Nonlinear regression)**.** Let the parameter space be the set of 1-Lipschitz functions

$$\mathscr{F} = \big\{\phi\colon [0,1] \to \mathbb{R}, \phi(0) = 0,\ |\phi(x) - \phi(x')| \le |x - x'| \text{ for all } x, x' \in [0,1]\big\},$$

equipped with the supremum metric $\rho_\infty(\theta, \theta') = \sup_x |\phi(x) - \phi'(x)|$. Suppose we observe $Y_i = \phi(x_i) + Z_i$, $i = 1, 2, \dots, n$, where $x_i \in [0,1]$ are fixed inputs and $Z_1, Z_2, \dots, Z_n$ are i.i.d. $\mathrm{N}(0, \sigma^2)$.

From Lemma 0.6 and Example 0.5, the logarithm of the $2\delta$-packing number of $\mathscr{F}$ under $\rho_\infty$ satisfies

$$\log M(2\delta; \mathscr{F}, \rho_\infty) \ge \log N(2\delta; \mathscr{F}, \rho_\infty) \ge c_1(1/2\delta)$$

for some $c_1 > 0$.

Let $f_i$ and $g_i$ denote the pdfs of $Y_i = \phi(x_i) + Z_i$ and $Y_i' = \phi'(x_i) + Z_i$, respectively. Their relative entropy is

$$D(f_i \| g_i) = \frac{\log e}{2\sigma^2} |\phi(x_i) - \phi'(x_i)|^2 \tag{0.264}$$

$$\leq \frac{\log e}{2\sigma^2} \sup_x |\phi(x) - \phi'(x)|^2 \tag{0.265}$$

$$= \Big(\rho_\infty(\phi, \phi')\sqrt{\log e/2\sigma^2}\Big)^2 \tag{0.266}$$

Define $\delta' = (\sigma\epsilon)\sqrt{\log e/2}$. Since from Example 0.5, the metric entropy of $\mathscr{F}$ at resolution $\delta'$ satisfies

$$\log N(\delta'; \mathscr{F}, \rho_\infty) \ \leq c_2(1/\delta')$$

for some $c_2 > 0$, which implies from (0.266) that the logarithm of the *relative entropy $\epsilon$-covering number* satisfies

$$\log K(\epsilon; \mathscr{P}) = \log N(\delta'; \mathscr{F}, \rho_\infty) \leq c_2(1/\delta') \leq c_3(1/\sigma\epsilon)$$

for some constant $c_3 > 0$.

Now, take $\epsilon = (1/\sigma n)^{1/3}$, $\delta = c(\sigma^2/n)^{1/3}$, and $c = c_1/4(c_3+2)$. Then,

$$\frac{\log K(\epsilon; \mathscr{P}) + n\epsilon^2 + 1}{\log M(2\delta; \Theta, \rho)} \leq \frac{c_3(1/\sigma\epsilon) + n\epsilon^2 + 1}{c_1/2\delta} \tag{0.267}$$

$$\leq \frac{c_3\sigma^{-2/3}n^{1/3} + \sigma^{-2/3}n^{1/3} + 1}{2(c_3+2)n^{1/3}\sigma^{-2/3}} \tag{0.268}$$

$$= \frac{c_3 + 1 + \sigma^{2/3}n^{-1/3}}{2(c_3+2)} \tag{0.269}$$

$$\leq \frac{1}{2} \tag{0.270}$$

for sufficiently large $n$. Substituting in (0.254), we obtain the minimax lower bound

$$\mathscr{R}_n(\mathscr{P}, \ell) \geq \frac{1}{2}\Phi\Big(c \cdot (\sigma^2/n)^{1/3}\Big). \tag{0.271}$$

## SUMMARY

**Covering and metric entropy.** Given a metric space $(\mathscr{V}, \rho)$ and $\delta > 0$, $\mathcal{V} = \{v_1, \ldots, v_m\} \subset \mathscr{V}$ is a *$\delta$-net* of $\mathscr{S} \subseteq \mathscr{V}$ if for every $v \in \mathscr{S}$, $\min_{1\leq i\leq m} \rho(v, v_i) \leq \delta$. The *covering number* $N(\delta; \mathscr{S}, \rho)$ of $\mathscr{S}$ is the size of its smallest $\delta$-net. The metric entropy of $\mathscr{S}$ is the logarithm of the covering number, $H(\delta; \mathscr{S}, \rho) = \log N(\delta; \mathscr{S}, \rho)$.

**Packing.** A *$\delta$-packing* of $\mathscr{S} \subseteq \mathscr{V}$ is a set of points $\{v_1, v_2, \ldots, v_m\} \subset \mathscr{S}$ such that $\rho(v_i, v_j) > \delta$ for every $i \neq j$. The packing number of $\mathscr{S}$, $M(\delta; \mathscr{S}, \rho)$, is the maximum $m$ such that there exists a $\delta$-packing with $m$ points.

**Rademacher complexity.** For $Z^n$ drawn i.i.d. from the data distribution and for i.i.d. Rademacher random variables $\epsilon_1, \ldots, \epsilon_n$ independent of $Z^n$, the expected Rademacher complexity of $\mathscr{W}$ is

$$\mathsf{R}_n(\mathscr{W}) = \mathsf{E}_{\varepsilon^n, Z^n}\Big(\frac{1}{n}\sup_{w\in\mathscr{W}}\Big|\sum_{i=1}^n \varepsilon_i \ell(Z_i, w)\Big|\Big).$$

**Worst-case generalization error bound.** For a learning problem with data space $\mathscr{Z}$, model class $\mathscr{W}$, and loss function $\ell : \mathscr{Z} \times \mathscr{W} \to \mathbb{R}_+$, $\Delta_n(\mathscr{W}) = \sup_{w\in\mathscr{W}} |L(w) - L_n(w)|$, where $L(w)$ is the expected loss with respect to the unknown data distribution and $L_n(w)$ is the empirical loss on an i.i.d. dataset.

1. $\mathsf{E}\left(\Delta_n(\mathscr{W})\right) \le 2\,\mathsf{R}_n(\mathscr{W})$.
2. A sufficient condition for generalization is $\mathsf{R}_n(\mathscr{W}) \le \sqrt{C(\mathscr{W})/n}$, where $C(\mathscr{W})$ is a quantity that captures the "generalization capacity" of $\mathscr{W}$.

**Expected generalization error bound.** For a learning algorithm $Z^n \to W$,

$$\left|\mathsf{E}\left(L(W) - L_n(W)\right)\right| \le \sqrt{\frac{2\sigma^2}{n} I(W; Z^n)}$$

provided $\ell(Z, w)$ is $\sigma^2$-subgaussian for every $w \in \mathscr{W}$.

**Gibbs algorithm.** $\mathscr{W} = \mathbb{R}^d$. Given a pdf $\pi(w)$ and inverse temperature $\beta > 0$,

$$f_\beta(w|z^n) = \frac{\exp\left(-\beta L_n(w)\right)\pi(w)}{\int_{\mathbb{R}^d} \exp\left(-\beta L_n(w')\right)\pi(w')\,dw'}.$$

Under appropriate regularity assumptions (including $\sigma^2$-subgaussianity)

$$\mathsf{E}(L(W)) - \inf_{w\in\mathscr{W}} L(w) \le O\Big(\frac{d}{\beta}\log\frac{\beta}{d}\Big) + \frac{\beta\sigma^2}{2n}.$$

**PAC-Bayes generalization bound.** Let $\ell(w, Z)$ be $\sigma^2$-subgaussian for every $w \in \mathscr{W}$. Then, for learning algorithm $f(w|z^n)$ and $\pi(w)$, with probability at least $1 - \delta$,

$$|L(W) - L_n(W)| \le \sqrt{\frac{2\sigma^2}{n-1}\Big(D(f(w|Z^n)\|\pi(w)) + \frac{1}{2}\ln n + \ln\frac{1}{\delta}\Big)}.$$

**Minimax estimation.** For a probability model $\mathscr{P}$ and loss function $\ell$, the minimax risk for estimating $\theta$ using i.i.d. data $\mathbf{X}_1, \mathbf{X}_2, \ldots, \mathbf{X}_n$ is

$$\mathscr{R}_n(\mathscr{P}, \ell) = \inf_{\hat\theta}\sup_{\theta} \mathsf{E}_{p_\theta}\left(\ell(\theta, \hat\theta(\mathbf{X}^n))\right).$$

**Fano-based bounds**. Let $\mathscr{P} = \{p_\theta : \theta \in \Theta\}$ be a probability model and $\ell$ be a loss function. Let $\{\theta_1, \theta_2, \ldots, \theta_m\}$ be any $2\delta$-packing of $\Theta$, and let $J \sim \mathrm{Unif}\{1, 2, \ldots, m\}$ and $\mathbf{X}^n|\{J = j\} \sim P_j^n$. Then, the minimax risk is lower bounded as

$$\mathscr{R}_n(\mathscr{P}, \ell) \ge \Phi(\delta)\Big(1 - \frac{I(J; \mathbf{X}^n) + 1}{\log m}\Big).$$

Local and global Fano methods provide order bounds on many estimation problems.

## PROBLEMS

**0.1.** *Bounds on Gaussian tails.* Let $Z \sim \mathrm{N}(0, \sigma^2)$. Prove the following upper and lower bounds on the tail probability $\Pr\{Z \geq a\}$ for $a > 0$:

$$\frac{\sigma}{a(1 + a^{-2})\sqrt{2\pi}} e^{-a^2/2\sigma^2} \leq \Pr\{Z \geq a\} \leq \frac{\sigma}{a\sqrt{2\pi}} e^{-a^2/2\sigma^2}.$$

Hint: For the upper bound, use the fact that the inequality $x/a \geq 1$ holds in the region of integration. For the lower bound, use the identity

$$\frac{1}{a} e^{-a^2/2} = -\int_a^\infty \frac{d}{dx}\left(\frac{1}{x} e^{-x^2/2}\right) dx.$$

**0.2.** *Scaling and adding subgaussian random variables.* Let $U_1$ and $U_2$ be zero-mean random variables, such that $U_i$ is $\sigma_i^2$-subgaussian, $i = 1, 2$.

(a) Show that $aU_1$ is $(a^2\sigma_1^2)$-subgaussian.

(b) Show that $U_1 + U_2$ is $(\sigma_1 + \sigma_2)^2$-subgaussian.
Hint: Use Hölder's inequality $|\mathsf{E}[(XY)]| \leq \left(\mathsf{E}|X|^p\right)^{1/p}\left(\mathsf{E}|Y|^q\right)^{1/q}$, where $p > 1$ and $1/p + 1/q = 1$, to upper-bound $\mathsf{E}\left(e^{\lambda(U_1+U_2)}\right)$. Then optimize over $p > 1$.

(c) Show that, when $U_1$ and $U_2$ are independent, the result of part (a) can be improved to: $U_1 + U_2$ is $(\sigma_1^2 + \sigma_2^2)$-subgaussian.

**0.3.** *Squared Gaussian random variable.* Let $Z \sim \mathrm{N}(0, \sigma^2)$. Show that, for any $0 \leq \lambda < 1/2\sigma^2$,

$$\mathsf{E}\left(e^{\lambda Z^2}\right) = \frac{1}{\sqrt{1 - 2\lambda\sigma^2}}.$$

**0.4.** *Proof of Hoeffding's lemma.* Prove the inequality (0.21). Hint: Use the fact that for any random variable $Z$ and $c \in \mathbb{R}$, $\mathrm{Var}(Z) \leq \mathsf{E}\left((Z - c)^2\right)$.

**0.5.** *Weak Hoeffding's lemma via symmetrization.* Let the random variable $X$ take values in $[a, b]$, then we can prove the bound

$$\mathsf{E}\left(e^{\lambda(X - \mathsf{E}(X))}\right) \leq \exp\left(\frac{\lambda^2(b-a)^2}{2}\right).$$

Justify the following steps.

(a) Let $X'$ be an independent copy of $X$, show that

$$\mathsf{E}\left(e^{\lambda(X - \mathsf{E}(X))}\right) \leq \mathsf{E}\left(e^{\lambda\varepsilon(X - X')}\right),$$

where $\varepsilon$ is a Rademacher random variable independent of $X, X'$.

(b) Use the result in Example 0.2 and the fact that $|X - X'| \leq b - a$ to establish the bound.

**0.6.** *Hoeffding's inequality and the AEP.* Let $X_1, X_2, \dots, X_n$ be independent random variables with $a_i \le X_i \le b_i$ and let $S_n = \frac{1}{n}\sum_{i=1}^n X_i$ and $\mu = \mathsf{E}(S_n)$. Combine Lemmas 0.1 and 0.2, to prove that for any $\epsilon > 0$,

$$\Pr\{|S_n - \mu| > \epsilon\} \le 2\exp\Big(-\frac{2n^2\epsilon^2}{\sum_{i=1}^n (b_i - a_i)^2}\Big). \tag{0.272}$$

This is known as Hoeffding's inequality. We can use it to provide the following "finite blocklength" AEP. Let $X_1, X_2, \dots$ be i.i.d. drawn from a finite set $\mathcal{X}$ with $p(x) > 0$ for every $x \in \mathcal{X}$. Let $a = \min_x -\log p(x)$ and $b = \max_x -\log p(x)$. Show that

$$\Pr\Big\{\Big|\sum_{i=1}^n -\log p_X(X_i) - nH(X)\Big| > n\epsilon\Big\} \le 2\exp\Big(-\frac{2n\epsilon^2}{(b-a)^2}\Big).$$

Finally, show that the probability of error $P(\mathcal{E})$ in the proof of achievability of the channel coding theorem decays exponentially in the block length $n$ if $R < C$. Hint: you already know that the second term in the bound on $P(\mathcal{E})$ decays exponentially in $n$.

**0.7.** *Pinsker inequality via Hoeffding's lemma.* Recall Pinsker's inequality,

$$\|p - q\|_{\mathrm{TV}}^2 \le \frac{\ln 2}{2} D(p\|q).$$

We can use the variational representation of relative entropy and Hoeffding's Lemma 0.1 to provide an alternative proof. Complete the following steps:

(a) Consider a set $\mathcal{A}$ such that $\|p - q\|_{\mathrm{TV}} = P(\mathcal{A}) - Q(\mathcal{A})$ and let $\phi_\lambda(x) = \lambda(\mathbf{1}_{\mathcal{A}}(x) - Q(\mathcal{A}))$. Verify that $\mathsf{E}_p(\phi_\lambda(X)) = \lambda\|p - q\|_{\mathrm{TV}}$ and $\mathsf{E}_q(\phi_\lambda(X)) = 0$.

(b) Substitute $\phi_\lambda$ in the variational formula for relative entropy.

(c) Noting that $\phi_\lambda(X)$ takes values in $[-Q(\mathcal{A}), 1 - Q(\mathcal{A})]$, use Hoeffding's lemma to show that

$$\log \mathsf{E}_q\Big(2^{\phi_\lambda(X)}\Big) \le \frac{\lambda^2 \ln 2}{8}.$$

(d) Complete the proof by combining steps (b) and (c) and maximizing the bound over $\lambda > 0$.

**0.8.** *Expected maximum of subgaussians.* Prove equation (0.36). Hint: Use the fact that $\max_{1\le i\le n} |X_i| = \max\{X_1, -X_1, \dots, X_n, -X_n\}$.

**0.9.** *Types as $\delta$-cover.* Consider the metric space consisting of the set $\mathscr{P}(\mathcal{X})$ of pmfs over a finite set $\mathcal{X}$ (i.e., probability simplex) and the total variation distance $\|p - q\|_{\mathrm{TV}}$. For what $\delta_n$ is the set of $n$-types $\mathcal{P}_n$ an $\delta_n$-net of $\mathscr{P}$?

**0.10.** *Metric entropy of a cube.* Consider the cube $[-1, 1]^d$ in $\mathbb{R}^n$ with the sup norm $\rho_\infty(\mathbf{v}, \mathbf{v}') = \max_{i\in\{1,2,\dots,n\}} |v_i - v_i'|$. Show that the covering number at resolution $\delta > 0$ is bounded as

$$H(\delta; [-1,1]^d, \rho_\infty, \delta) \le n \log\Big(1 + \frac{1}{\delta}\Big).$$

**0.11.** *Upper bound on metric entropy of Lipschitz functions.* Show that for the functions in Example 0.5, $H(\delta; \mathscr{F}, \rho_\infty) \le \lceil 2L/\delta \rceil \log 3$. Hint: To prove the upper bound, assume that $0 < \delta < L/2$ and partition the $[0, 1]$ interval into $m = \lceil 2L/\delta \rceil$ equal size subintervals of length $\Delta = 1/m$. For each $\beta = (\beta_1, \beta_2, \ldots, \beta_m) \in \{-1, 0, 1\}^m$, define $\phi_\beta$ to be a piecewise linear function with $\phi_\beta(0) = 0$ and $d\phi_\beta(v)/dv = \beta_j L$ on subinterval $j = 1, 2, \ldots, m$. Show that this set of functions constitutes a $\delta$-cover of $\mathscr{F}$.

**0.12.** *Packing.* Prove Lemma 0.6, which states that for any $\delta > 0$, the covering and packing numbers for a set $\mathscr{S}$ at resolution $\delta$ are related as

$$M(2\delta; \mathscr{S}, \rho) \le N(\delta; \mathscr{S}, \rho) \le M(\delta; \mathscr{S}, \rho).$$

(a) Use the above result to show that for the Hamming cube $\{0, 1\}^n$, assuming $\delta < n/2$,

$$\frac{2^n}{\sum_{i=0}^{2\delta} \binom{n}{i}} \le M(\delta; \{0, 1\}^n, \rho_{\mathrm{H}}) \le \frac{2^n}{\sum_{j=0}^{\delta} \binom{n}{j}}.$$

Note the similarity between these bounds and the bounds on the size of an $(n, k)$ error correcting code with minimum distance $2\delta$.

(b) For $\delta = np$, $0 < p < 1/2$, use the bounds in part (b) to show that

$$1 - \mathsf{H}(2p) \le \frac{1}{n} \log M(np; \{0, 1\}^n, \rho_{\mathrm{H}}) \le 1 - \mathsf{H}(p).$$

Comment on the relationship to the capacity of the binary symmetric channel.

**0.13.** *Metric entropy lower bound for Hamming cube.* Show that the lower bound in Example 0.4 can be loosened to

$$H\big(\delta; \{0, 1\}^n, \rho_{\mathrm{H}}\big) \ge nD\big(\mathrm{Bern}(\delta/n) \big\| \mathrm{Bern}(1/2)\big).$$

Hint: Let $X_1, X_2, \ldots, X_n$ be i.i.d. Bern(1/2). Show that

$$2^{-n} V(n, \delta) = \mathsf{Pr}\Big\{ \sum_i X_i \le \delta \Big\} \le 2^{-nD\left(\mathrm{Bern}(\delta/n) \| \mathrm{Bern}(1/2)\right)}.$$

**0.14.** *Infinite model class.* Let $\mathscr{Z} = \mathscr{X} \times \mathscr{Y}$, where $\mathscr{X}$ is a compact subset of $\mathbb{R}^d$ and $\mathscr{Y} \subseteq \mathbb{R}$. Let $\phi(\cdot, w) : \mathscr{X} \to \mathscr{Y}$ be a given family of functions, parametrized by the elements $w$ of a totally bounded metric space $(\mathscr{W}, \rho)$. For $Z = (X, Y)$, we can think of $\phi(X, w)$ as a candidate predictor of $Y$. Suppose that the following inequality holds for some constant $c > 0$ for all $x \in \mathcal{X}$ and all $w, w' \in \mathscr{W}$: $|\phi(x, w) - \phi(x, w')| \le c\rho(w, w')$. Consider the absolute loss given by $\ell(z, w) = \ell((x, y), w) = |y - \phi(x, w)|$. Verify that (0.72) holds for any $w, w' \in \mathscr{W}$. Show that for any outcome $z^n$, $|L_n(w) - L_n(w')| \le c\rho(w, w')$.

**0.15.** *VC classes and metric entropy.* Let $\mathscr{C}$ be a VC class of measurable subsets of $\mathscr{X}$, and let $D$ denote its VC dimension. Let $Q$ be a probability measure on $\mathscr{X}$.

(a) Let $\mathscr{A}_1, \ldots, \mathscr{A}_N$ be a $\delta$-packing of $\mathscr{C}$ with respect to the metric $\rho_Q(\mathscr{A}, \mathscr{B}) = \sqrt{Q(\mathscr{A} \Delta \mathscr{B})}$. Use a random coding argument to show that there exist $M = \lceil (2/\delta^2) \log N \rceil + 1$ points $x_1, \ldots, x_M \in \mathscr{X}$, such that $\{x_1, \ldots, x_M\} \cap (\mathscr{A}_i \Delta \mathscr{A}_j) \neq \emptyset$ for all $1 \leq i, j \leq N$ with $i \neq j$. Hint: Let $X_1, \ldots, X_M$ be i.i.d. draws from $Q$. Use the $\delta$-packing property of $\mathscr{A}_1, \ldots, \mathscr{A}_N$ to show that, for the above choice of $M$,

$$\Pr\left(\{X_1, \ldots, X_M\} \cap (\mathscr{A}_i \Delta \mathscr{A}_j) \neq \emptyset \text{ for all } i \neq j\right) > 0.$$

(b) Show that the finite class of sets $\hat{\mathscr{C}} = \{\mathscr{A}_1, \ldots, \mathscr{A}_N\}$ cannot shatter any subset of $\mathscr{X}$ of size larger than $D$, i.e., for any $n \geq D$,

$$\sup_{x_1, \ldots, x_n \in \mathscr{X}} \left| \left\{ \left( \mathbf{1}_{\{x_1 \in \mathscr{A}_i\}}, \ldots, \mathbf{1}_{\{x_n \in \mathscr{A}_i\}} \right) : 1 \leq i \leq N \right\} \right| < 2^n.$$

Use this together with Lemma 0.11 to show that, if $M \geq D$, then

$$N \leq \left( \frac{eM}{D} \right)^D.$$

(c) Show that, if $\mathscr{A}_1, \ldots, \mathscr{A}_N$ is a $\delta$-packing of $\mathscr{C}$ w.r.t. the metric $\rho_Q$, then

$$\log N \leq 2D \log \frac{4e}{\delta^2}.$$

Hint: Since $\delta \in [0, 1]$, you only need to consider the case $\log N \geq D$, which is covered by part (b).

(d) Finally, prove the upper bound (0.134) for the metric entropy of $\mathscr{C}$ with respect to the metric $\rho_Q$.

**0.16.** *Generalization bound using individual mutual information.* Consider the setting of Theorem 0.6.

(a) Under the same assumptions, prove the following generalization bound involving the mutual information between the algorithm's output $W$ and each individual data sample $Z_i$:

$$\left| \mathsf{E}\left( L(W) - L_n(W) \right) \right| \leq \frac{1}{n} \sum_{i=1}^{n} \sqrt{2\sigma^2 I(W; Z_i)}.$$

(b) Show that the bound from part (a) is tighter than the bound of Theorem 0.6.

(c) Consider the case when the $Z_i$s are i.i.d. $\mathrm{N}(\theta, \sigma^2)$ random variables and let $W = \frac{1}{n} \sum_i Z_i$. Show that, in this case, the mutual information $I(W; Z^n)$ is infinite, while the quantity $\frac{1}{n} \sum_{i=1}^{n} \sqrt{2\sigma^2 I(W; Z_i)}$ is finite and evaluate it explicitly.

**0.17.** *The ERM limit of the Gibbs algorithm.* Consider the Gibbs algorithm given by the conditional pdf in (0.158) and assume the following: $\mathscr{W} = \mathbb{R}^d$, $\pi(w) = \mathrm{N}(w; 0, I_d)$, and the loss function $\mathcal{L}$ is smooth in the sense that there exists a constant $M > 0$ such that, for any dataset $z^n$,

$$\left|\mathcal{L}(w, z^n) - \min_w \mathcal{L}(w, z^n)\right| \le \frac{M}{2}\|w - w^*\|^2 \qquad \text{for every } w \in \mathbb{R}^d,$$

where $w^*$ is the unique minimizer of $\mathcal{L}(w, z^n)$ over $w$. For an i.i.d. dataset $Z^n$ and $\beta > 0$, let $W_\beta$ be the random output of the Gibbs algorithm.

(a) Show that, for any $\beta > 0$,

$$L_n(W_\beta) \le -\frac{1}{\beta}\log c_\beta(Z^n),$$

where

$$c_\beta(Z^n) = \mathsf{E}_{\tilde{W}\sim\pi}\left(e^{-\beta\mathcal{L}(\tilde{W}, Z^n)}\right).$$

Hint: Compute the relative entropy $D(f_\beta(w|Z^n)\|\pi(w))$.

(b) Next, show that

$$\log c_\beta(Z^n) \ge -\beta\inf_w \mathcal{L}(w, Z^n) + \log\int_{\mathbb{R}^d} e^{-\frac{\beta M}{2}\|w-w^*\|^2}\pi(w)\,dw. \tag{0.273}$$

(c) We now further lower bound the integral on the right-hand side of (0.273). For $\epsilon > 0$, to be chosen later, lower bound the integral by restricting the region of integration to the ball of radius $\epsilon$ centered at $w^*$ and conclude that

$$\int_{\mathbb{R}^d} e^{-\frac{\beta M}{2}\|w-w^*\|^2}\pi(w)\,dw \ge \exp\left(-\frac{1}{2}(\epsilon + \|w^*\|^2)\right)\cdot\frac{\epsilon^d V_d}{(2\pi)^{d/2}},$$

where $V_d$ is the volume of the Euclidean unit ball in $\mathbb{R}^d$.

(d) Combine the above steps and show that, for an appropriate choice of $\epsilon$, the empirical loss of $W_\beta$ on $Z_1, Z_2, \ldots, Z_n$ satisfies

$$L_n(W_\beta) \le \inf_{w\in\mathbb{R}^d} L_n(w) + \frac{d}{\beta}\log\frac{A\beta}{d},$$

where $A > 0$ is a quantity that depends on $\|w^*\|$ and on the smoothness constant $M$. Conclude that the Gibbs algorithm converges to ERM as $\beta \to \infty$.

**0.18.** *Minimax estimation of Gaussian vector mean.* Extend Example 0.8 to minimax estimation of the mean $\theta^d$ of Gaussian vector $X^d \sim \mathrm{N}(\theta^d, \sigma^2 I)$.

**0.19.** *Hellinger versus relative entropy* Show that for pdfs $f_1, f_2 : [0,1] \to [c_1, c_2]$, $c_1 < c_2$,

$$D(f_1\|f_2) \le \frac{1}{c_1\ln 2}D_{\mathrm{H}}^2(f_1, f_2).$$

## BIBLIOGRAPHIC NOTES

The book by Boucheron, Lugosi, and Massart (2013) is an excellent source on concentration inequalities and related matters. The concept of metric entropy has its origins in a seminal paper by Kolmogorov and Tikhomirov (1959). The relationship between metric entropy and rate–distortion is discussed in the book by Polyanskiy and Wu (2025), where it is shown that, in general, metric entropy can be upper and lower bounded in terms of the "worst-case" rate–distortion function over the source distribution.

The framework of statistical learning theory used in this chapter was developed by Vapnik (1982) and (1991). From the perspective of computational complexity, the notion of PAC learning was first formalized by Valiant (1984) and Angluin (1988). The connection to statistical learning theory was subsequently developed by Haussler (1992).

The central importance of the worst-case generalization error in the context of Empirical Risk Minimization was first recognized by Vapnik and Chervonenkis (1971), (1981). Their analysis of it made use of symmetrization in the form of introducing an independent copy $\bar{Z}^n$ of the training data $Z^n$ and then applying a permutation chosen uniformly at random to the entire sequence of length $2n$. The first simplification of their original proof using Rademacher averages was due to Koltchinskii (1981) and subsequently by Giné and Zinn (1984). The definition and analysis of what is now called the VC dimension also appeared in the original works of Vapnik and Chervonenkis.

Information-theoretic bounds on the expected generalization error were obtained by Russo and Zou (2016), Raginsky, Rakhlin, Tsao, Wu, and Xu (2016), and Xu and Raginsky (2017). The bound using individual mutual information is due to Bu, Zou, and Veeravalli (2019); also see the book chapter by Raginsky, Rakhlin, and Xu (2021). PAC-Bayes formulation of statistical learning was introduced by McAllester (1999); also see the paper by Zhang (2006) and the monograph by Catoni (2007). The monograph by Hellström, Durisi, Guedj, and Raginsky (2025) presents generalization bounds from complementary information-theoretic and PAC-Bayes perspectives.

Classical generalization bounds, such as those based on VC dimension and Rademacher complexity, measure the capacity of a model class. As these measures typically scale with the number of parameters or matrix norms, they predict that highly overparameterized neural networks should severely overfit. However, this prediction is starkly contradicted by empirical results. As demonstrated by Zhang, Bengio, Hardt, Recht, and Vinyals (2017), modern deep networks can achieve zero training error even on datasets with random labels—a clear sign of their enormous capacity—yet still generalize well when trained on real data. This phenomenon renders classical, data-independent capacity measures vacuous in realistic deep learning settings. This mismatch has catalyzed a shift toward more refined analyses that are dependent on the training data and the learning algorithm; see for example, (Bartlett, Montanari, and Rakhlin 2021).

The minimax estimation framework is due to Wald (1950). Lower bounds using Fano's inequality first appeared in (Ibragimov and Hasminskii 1977) and (Hasminskii 1978); also see the book by Ibragimov and Hasminskii (1981). The global Fano method is due to Yang and Barron (1999). The book by Wainwright (2019) provides a detailed coverage of mini-max bounds, including using the Le Cam (1973) and Assouad (1983) methods, in addition to the Fano methods discussed in this chapter.

Although we do not discuss lower bounds on the performance of learning algorithms in this chapter, they are typically stated in terms of the minimax excess risk and derived using the information-theoretic tools used to bound minimax estimation risk. The resulting bounds scale with measures of model class complexity and characterize the fundamental order-wise limits of performance across a range of problems in classification, regression, and nonparametric learning; see, e.g.,(Tsybakov 2009, Wainwright 2019).

# Bibliography

Angluin, D. (1988). Queries and concept learning. *Machine Learning*, 2(4), 319–342. [57]

Assouad, P. (1983). Deux remarques sur l'estimation. *Comptes Rendus de l'Académie des Sciences, Paris*, 296, 1021–1024. [58]

Bartlett, P. L., Montanari, A., and Rakhlin, A. (2021). Deep learning: a statistical viewpoint. *Acta Numerica*, 30, 87–201. [57]

Bottou, L. and Bousquet, O. (2007). The tradeoffs of large scale learning. In *Advances in Neural Information Processing Systems (NeurIPS)*, vol. 20. [16]

Boucheron, S., Lugosi, G., and Massart, P. (2013). *Concentration Inequalities: A Nonasymptotic Theory of Independence*. Oxford University Press. [25, 57]

Bu, Y., Zou, S., and Veeravalli, V. V. (2019). Tightening mutual information based bounds on generalization error. In *2019 IEEE International Symposium on Information Theory (ISIT)*, pp. 587–591. [57]

Catoni, O. (2007). *PAC-Bayesian supervised classification: The thermodynamics of statistical learning*, vol. 56 of *IMS Lecture Notes–Monograph Series*. Institute of Mathematical Statistics. [57]

Dudley, R. M. (1978). Central limit theorems for empirical measures. *The Annals of Probability*, 6(6), 829–929. [27]

Giné, E. and Zinn, J. (1984). Some limit theorems for empirical processes. *The Annals of Probability*, 12(4), 928–989. [57]

Giraud, C. (2021). *Introduction to High-Dimensional Statistics*. 2nd ed. Chapman & Hall/CRC, Boca Raton, FL, USA. [37]

Hasminskii, R. Z. (1978). A lower bound on the risks of nonparametric estimates of densities in the uniform metric. *Theory of Probability and Its Applications*, 23, 794–798. [57]

Haussler, D. (1992). Decision theoretic generalizations of the PAC model for neural net and other learning applications. *Information and Computation*, 100(1), 78–150. [13, 57]

Hellström, F., Durisi, G., Guedj, B., and Raginsky, M. (2025). Generalization bounds: Perspectives from information theory and PAC-Bayes. *Foundations and Trends in Machine Learning*, 18(1), 1–223. [57]

Ibragimov, I. A. and Hasminskii, R. Z. (1977). On the estimation of an infinite-dimensional parameter in gaussian white noise. *Soviet Mathematics Doklady*, 18, 1307–1309. [57]

Ibragimov, I. A. and Hasminskii, R. Z. (1981). *Statistical Estimation: Asymptotic Theory*. Springer, New York, NY. [57]

Kolmogorov, A. N. and Tikhomirov, V. M. (1959). $\epsilon$-entropy and $\epsilon$-capacity of sets in functional spaces. *Uspekhi Matematicheskikh Nauk*, 14(2), 3–86. English translation: *Amer. Math. Soc. Transl. Ser. 2*, 17 (1961), 277–364. [7, 57]

Koltchinskii, V. I. (1981). On the central limit theorem for empirical measures. *Probab. Theory Math. Statist.*, 24(79-82). [57]

Le Cam, L. (1973). Convergence of estimates under dimensionality restrictions. *The Annals of Statistics*, 1(1), 38–53. [58]

Ledoux, M. and Talagrand, M. (1991). *Probability in Banach Spaces*. Springer Berlin Heidelberg. [20]

McAllester, D. A. (1999). Some PAC-Bayesian theorems. *Machine Learning*, 37(3), 355–363. [16, 57]

Polyanskiy, Y. and Wu, Y. (2025). *Information Theory: From Coding to Learning*. Cambridge University Press. [57]

Raginsky, M., Rakhlin, A., Tsao, M., Wu, Y., and Xu, A. (2016). Information-theoretic analysis of stability and bias of learning algorithms. In *2016 IEEE Information Theory Workshop (ITW)*, pp. 26–30. [57]

Raginsky, M., Rakhlin, A., and Xu, A. (2021). Information-theoretic stability and generalization. In M. R. D. Rodrigues and Y. C. Eldar (eds.) *Information-Theoretic Methods in Data Science*, chap. 10, pp. 302–329. Cambridge University Press. [57]

Russo, D. and Zou, J. (2016). Controlling bias in adaptive data analysis using information theory. In *Proceedings of the 19th International Conference on Artificial Intelligence and Statistics*, vol. 51, pp. 1232–1240. [57]

Sauer, N. (1972). On the density of families of sets. *Journal of Combinatorial Theory, Series A*, 13(1), 145–147. [27]

Shelah, S. (1972). A combinatorial problem; stability and order for models and theories in infinitary languages. *Pacific Journal of Mathematics*, 41(1), 247–261. [27]

Tsybakov, A. (2009). *Introduction to Nonparametric Estimation*. Springer Series in Statistics. Springer, New York. [37, 58]

Valiant, L. (1984). A theory of the learnable. *Commun. ACM*, 1134–1142. [13, 57]

Vapnik, V. (1982). *Estimation of Dependencies Based on Empirical Data*. Springer-Verlag, New York. [57]

Vapnik, V. (1991). *The Nature of Statistical Learning Theory*. Springer-Verlag, New York. [57]

Vapnik, V. and Chervonenkis, A. (1971). On the uniform convergence of relative frequencies to their probabilities. *Theory Prob. Appl.*, 264–280. [27, 57]

Vapnik, V. and Chervonenkis, A. (1981). Necessary and sufficient conditions for the uniform convergence of means to their expectations. *Theory Prob. Appl.*, 532–553. [57]

Verdú, S. and Weissman, T. (2008). The information lost in erasures. *IEEE Transactions on Information Theory*, 54(11), 5030–5058. [31]

Wainwright, M. J. (2019). *High-Dimensional Statistics: A Non-Asymptotic Viewpoint*. Cambridge University Press. [46, 57, 58]

Wald, A. (1950). *Statistical Decision Functions*. Wiley Publications in Statistics. John Wiley & Sons, New York. Reprinted in *Breakthroughs in Statistics*, Springer, 1992. [37, 57]

Xu, A. and Raginsky, M. (2017). Information-theoretic analysis of generalization capability of learning algorithms. In *Advances in Neural Information Processing Systems (NeurIPS)*, vol. 30. [57]

Yang, Y. and Barron, A. R. (1999). Information-theoretic determination of minimax rates of convergence. *The Annals of Statistics*, 27(5), 1564–1599. [57]

Zhang, C., Bengio, S., Hardt, M., Recht, B., and Vinyals, O. (2017). Understanding deep learning requires rethinking generalization. In *International Conference on Learning Representations (ICLR)*. [57]

Zhang, T. (2006). Information-theoretic upper and lower bounds for statistical estimation. *IEEE Transactions on Information Theory*, 52(4), 1307–1321. [57]